\documentclass{aa}  

\usepackage{graphicx}
\usepackage{color}
\usepackage{txfonts,amsmath}
\usepackage{hyperref}
\usepackage{booktabs}
\usepackage{multirow}

\newcommand{\T}[1]{$\Delta \theta\ = #1''$}
\renewcommand{\d}[1]{$d = #1\ \text{kpc}$}
\newcommand{\z}[1]{$z = #1$}

\newcommand{\LIS}{Legacy Survey Imaging}

\begin{document}

   \title{Discovery of 19 strongly-lensed quasars, dual and projected quasars in DESI-LS}

   \author{Zizhao He
          \inst{1}
          \and
          Qihang Chen\inst{2,3}
          \and
          Limeng Deng \inst{1,4}
          \and
          Yiping Shu\inst{1}
          \and
          Rui Li\inst{5}
          \and
          Nan Li\inst{6}
          \and
          Dongdong Shi\inst{7}
          \and
          Guoliang Li\inst{1}
          }

   \institute{Purple Mountain Observatory, Chinese Academy of Sciences, Nanjing, Jiangsu, 210023, China\\
              \email{zzhe@pmo.ac.cn}
        \and
            School of Physics and Astronomy, 
            Beijing Normal University, Beijing, 100875, China
         \and
            Institute for Frontier in Astronomy and Astrophysics, 
            Beijing Normal University, Beijing, 102206, China
        \and
            School of Astronomy and Space Sciences, 
            University of Science and Technology of China, Hefei 230026, China
        \and
            Institute for Astrophysics, School of Physics, Zhengzhou University, Zhengzhou, 450001, China
        \and
            Key lab of Space Astronomy and Technology, National Astronomical Observatories, 20A Datun Road, Chaoyang District, Beijing 100012, China; 
        \and
            Centre for Fundamental Physics, School of Mechanics and Optoelectric Physics, Anhui University of Science and Technology, Huainan, Anhui 232001, China\\
             }

   \date{Received September 15, 1996; accepted March 16, 1997}

  \abstract
    {We report the follow-up spectroscopic confirmation of two lensed quasars, six dual quasars, and eleven projected quasars that were previously identified as lensed-quasar candidates in \cite{He2023}. The spectroscopic data were obtained from two different sources: the P200/DBSP in California and publicly available datasets, including SDSS and DESI-EDR. The two lensed quasars (both pairs) have the following properties: $\theta_E$ = 1$\arcsec$.208, $z_s$ = 3.105; $\theta_E$ = 0$\arcsec$.749, $z_s$ = 2.395. The six dual quasars have redshifts ranging from 0.58 to 3.28 and projected separations ranging from 15.44 to 22.54 kpc, with a mean separation of 17.95 kpc. The eleven projected quasars have projected separations ranging from 10.96 to 39.07 kpc, with a mean separation of 22.64 kpc. Additionally, there are three likely lensed quasars that cannot be definitively confirmed, contributed by two reasons. Firstly, their image separations (0$\arcsec$.83, 0$\arcsec$.98, and 0$\arcsec$.93) are small compared to the seeing conditions during our observations (around 1$\arcsec$.2). Secondly, no high SNR lensing galaxy can be detected in the \LIS. Better spectroscopy and (or) imaging are needed to confirm their lensing nature.}

   \maketitle
%

\section{Introduction}

When two quasars with a small separation are observed, there are often three different physical scenarios. The most common scenario is projected quasars, where quasars coincidentally appear very close to each other along the line of sight, but are actually at vastly different redshifts. A rarer scenario is dual quasars, which are at similar redshift and physically interacting. The rarest scenario is lensed quasars, where the light from a single quasar is bent, resulting in two images of the same quasar. In that case, an intervening lensing galaxy is usually expected.

The three categories mentioned above each play different roles in the investigation of the universe. Gravitationally lensed quasars not only enable investigations traditionally addressed by galaxy-galaxy lenses, such as constraining the equation of state of dark energy and investigating galaxy evolution \citep{Oguri2014,Suyu2014,Shu2015,Sonnenfeld2021,Filipp2023}, but also provide a unique avenue for measuring the Hubble constant \citep{Shajib2018,Liao2019,Wong2020,Li2021}. Moreover, they offer information on the structures of active galactic nuclei (AGNs) \citep{Anguita2008, Sluse2011, Motta2012, Guerras2013, Braibant2014, Fian2021, Hutsemekers2021}. Dual quasars, which refer to quasar pairs separated by 1 pc to 100 kpc \citep{DeRosa2019}, have considerable significance in elucidating the growth and evolution of binary supermassive black holes \citep[BSMBHs,][]{Roedig2014,Romero2016} and illuminating the processes of galactic-scale merging \citep{BoylanKolchin2008,Martin2018}. Projected quasars can be used to study the circumgalactic medium \citep[CGM,][]{Cai2019} properties of quasar host galaxies \citep{Findlay2018,Chen2023}.

The discovery of those objects involves two key stages: candidate selection followed by spectroscopy confirmation. The first often rely on imaging surveys such as Dark Energy Spectroscopic Instrument Legacy Imaging Surveys \citep[DESI-LS,][]{Dey2019}, Kilo Degree Survey \citep[KiDS,][]{deJong2019}, Dark Energy Survey \citep[DES,][]{DES2005}, Panoramic Survey Telescope and Rapid Response System \citep[Pan-STARRS,][]{Panstarss2016}, or astrometric satellites such as Gaia \citep{GaiaCollaboration2018}. The second relies on spectroscopic surveys, such as the Sloan Digital Sky Survey \citep[SDSS,][]{Blanton2017} and the Dark Energy Spectroscopic Instrument \citep[DESI,][]{DESICollaboration2016}, and targeted follow-up observations \citep[e.g.,][]{lemon2022,Dux2023,Dux2024}.

Based on random forest selection and DESI-LS data, 620 new candidate lensed quasars have been collected in \cite{He2022,He2023}. In this study, we selected ten high-priority observable candidates from \cite{He2023} and obtained their spectra using the DBSP/P200 at Palomar Observatory. Most of these targets exhibit separations smaller than $2.4\arcsec$. Prior to our observations, we checked these targets in recently available spectral datasets, including SDSS DR16 and DESI EDR, to verify whether they had been spectroscopically observed in recent surveys. For confirmed lensed quasars, we performed light and lens modelling to reveal their properties. For confirmed dual quasars, we calculated velocity differences, determined projected separations, and discussed CGM features.

The structure of this paper is as follows.
Section \ref{sec:obs} details the observational aspects, including a review of the target selection process, spectral data reduction procedures, and specifics of spectroscopic follow-up. Sections \ref{sec:p200res} and \ref{sec:specdataset} present the confirmation results of P200/DBSP observations and publicly available datasets, respectively. Finally, Section \ref{sec:diss} provides a comprehensive discussion and summary of our findings. In this paper, a fiducial cosmological model with $\Omega_m=0.26$, $\Omega_{DE}=0.74$, $h=0.72$, $w_0=-1$, and $w_a=0$ is assumed. Unless otherwise stated, all magnitudes quoted in this paper are in the AB system.

\section{Observations}\label{sec:obs}

In Section \ref{subsec:target_sel}, we briefly review the target selection process, with further details available in \cite{He2022,He2023}. Section \ref{subsec:spec_setup} introduces the spectroscopic equipment setups and observing conditions. Finally, Section \ref{subsec:spec_reduction} provides a brief overview of the spectral data reduction process.

\subsection{Target selection} \label{subsec:target_sel}

The first step in our approach involves employing a Random Forest (RF) classifier to identify quasar candidates from the DESI Legacy Surveys (DESI-LS) photometry catalogue. Our training set comprises 651,073 positive and 1,227,172 negative samples, incorporating photometric data from both DESI-LS and the Wide-field Infrared Survey Explorer (WISE). The labels for these samples were derived from the Sloan Digital Sky Survey (SDSS) and the Set of Identifications, Measurements, and Bibliography for Astronomical Data (SIMBAD).

We applied the trained RF model to point-like sources in DESI-LS Data Release 9. The performance of the classifier, evaluated using a test set, demonstrated a recall of $\sim$99\% at a purity of $\sim$30\%. Through this process, we successfully identified approximately 24 million quasar candidates from a pool of $\sim$0.42 billion point-like sources.
In the second phase, we designed a quasar group finder algorithm based on spatial coordinates to identify quasars in close proximity to each other. This algorithm initially generated $\sim$560,000 quasar groups from the $\sim$24 million quasar candidates. We then refined these groups by analysing the similarity of colours among group members and their likelihood of being quasars. This refinement process reduced the number of quasar groups from $\sim$560,000 to $\sim$140,000.

The final stage involved visual inspection to select candidate strongly lensed quasars based on the spatial configuration of group members. During this process, we classify candidates into categories A, B, and C, with grade A representing the most promising candidates. This examination yielded 620 new lensed quasar candidates, consisting of 101 grade-A, 214 grade-B, and 305 grade-C candidates. Following the name used in \cite{He2023}, we refer to the catalogue as H22.

\subsection{Spectroscopic follow-up}\label{subsec:spec_setup}

The selection of targets followed a two-step process. The criteria of the first step were:
\begin{itemize}
    \item Magnitudes of quasars are brighter than 21.5 in $r$-band;
    \item No prior spectroscopic observations reported;
    \item Classified as grade A or B in H22;
    \item Observable in Palomar observatory on October 15-16, 2023.
\end{itemize}
Subsequently, two additional conditions were considered, and it is sufficient for one of them to be met:
\begin{itemize}
    \item A extend component can be identified between two point-like sources in the residuals of \LIS;
    \item Also flagged as candidates in \cite{Dawes2023}.
\end{itemize}
After applying these criteria, 21 targets were selected, of which 10 were observed due to weather constraints.

Long-slit spectroscopic follow-up was carried out for those ten candidates using the Double Spectrograph (DBSP) equipped on the P200 on the nights of 15-16 Oct. 2023 (P.I. Zizhao He). Their magnitudes and separations can be found in Table \ref{tab:obs}. 

\begin{figure*}
    \centering
    \includegraphics[width=1\textwidth]{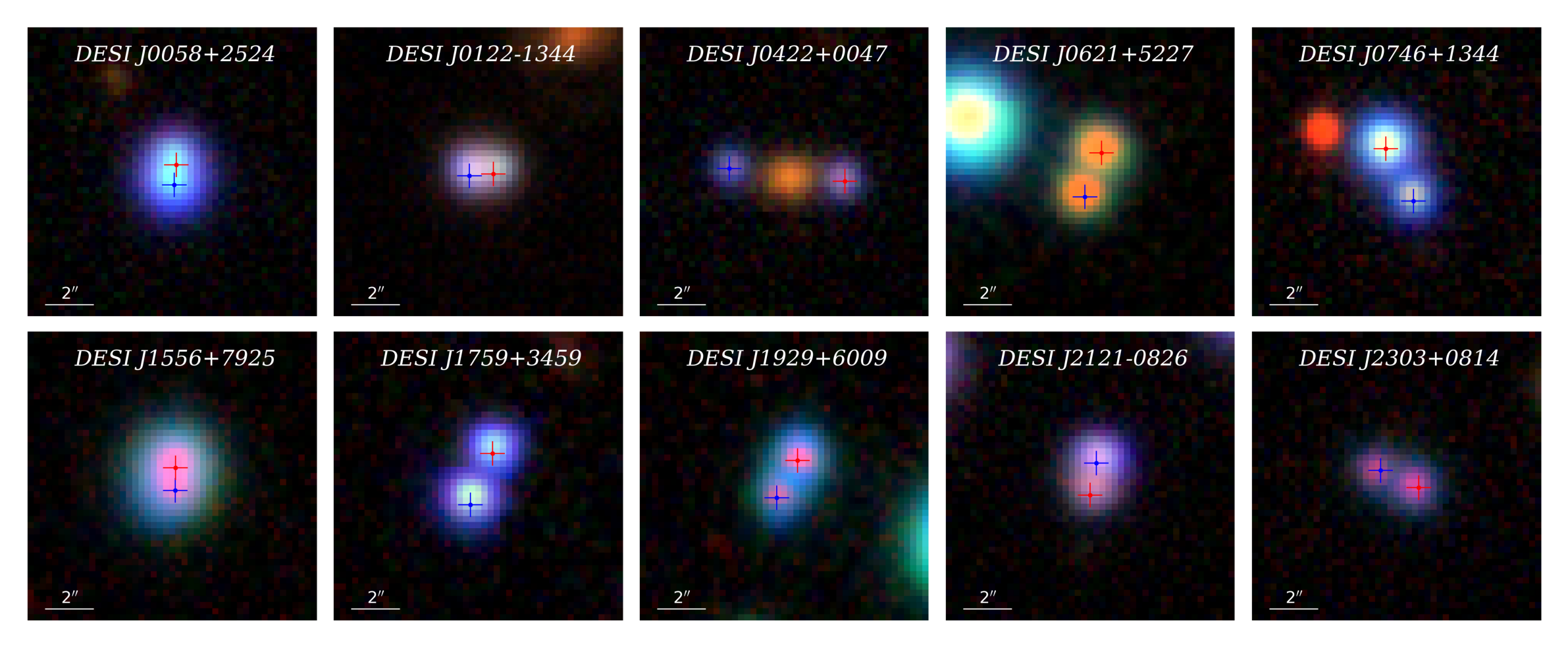}
    \caption{DESI-LS colour images of the 10 targets observed by DBSP/P200 during Oct 15-16, 2023. The cutout size of each image is $12^{\prime\prime} \times 12^{\prime\prime}$ and north is up while east is left. These cutout images are obtained from \LIS\ DR10 in \textit{grz} bands and re-plotted in this work using {\tt HumVI} \citep{Marshall2016}.}
    \label{fig:obs}
\end{figure*}

The brightest magnitude is $r=18.45$ and the faintest is $r=21.14$, with a median value of $r=19.60$. The largest separation is 4$\arcsec$.65 while the minimum is 0$\arcsec$.83, with a median value of 1$\arcsec$.78. The composite image of the $grz$ bands can be found in Figure \ref{fig:obs}. The standard dichroic D-55 was used to split light into blue and red ends. The 300 lines/mm grating blazed at 3990 \AA\ was chosen for blue arms, while the 316 lines/mm grating blazed at 7150 \AA for red arms. This provides a dispersion of 2.108 \AA/pixel and 1.535 \AA/pixel for the blue and red arms \citep{Oke1982}. 

Seeing conditions generally ranged between 0.9$\arcsec$ and 1.5$\arcsec$, and a 1.5$\arcsec$ wide slit was utilized. For each exposure, the slit was aligned with two point sources, ensuring the sources were as close to the centre of the slit as possible. This approach ensured simultaneous and efficient acquisition of spectra for both sources.

Table \ref{tab:obs} presents the key observational characteristics of the targets, including coordinates, exposure times, image separations, LS-DR10 $r$-band magnitudes, and the determined redshifts. In addition, the table provides the results of the spectroscopic identifications.

\subsection{Spectral data reduction}\label{subsec:spec_reduction}

The raw spectra were reduced using a \textsc{Python}\footnote{\url{https://www.python.org/}} code developed by us. The code utilizes several well-known Python modules, including \texttt{Astropy} \citep{AstropyCollaboration2022}, \texttt{NumPy} \citep{Harris2020}, and \texttt{Pandas} \citep{McKinney2010}. We implemented a standard data reduction procedure that includes bias subtraction, flat-fielding, cosmic ray rejection \citep[\texttt{Astroscrappy}, ][]{McCully2018}, 1D spectrum extraction, and wavelength and flux calibrations. 

Wavelength calibration at the blue and red ends was performed using FeAr and HeNeAr blended arc lamps, respectively. Flux calibration for the blue end was carried out with standard stars Feige 15 and Feige 34, while the red end was calibrated using Feige 34. Furthermore, all reduced spectra, along with those from SDSS DR16 and DESI EDR, were smoothed using Gaussian filtering to reduce noise and enhance spectral features.

\begin{table*}[]
    \caption{Observing table of 10 objects. `H22-ID' is the ID of the system in \cite{He2023}.}
    \tiny
    \centering
    \begin{tabular}{llccccclc}
    \hline
    Name           & H22-ID   & RA($^\circ$)       & Dec($^\circ$)        & Exp. Time (s) & Separation ($^{\prime\prime}$) & r$_{mag}$        & Quasar Redshift & Outcome     \\ \hline
    DESI J0058+2524 & 2459355  & 14.539675  & +25.409982 & 1600 & 0.83 & 18.94, 19.71 & 2.580 & likely lensed quasars        \\
    DESI J0122-1344 & 10930010 & 20.656958  & -13.739587 & 2000 & 0.98 & 19.87, 20.26 & 1.420 & likely lensed quasars \\ 
    DESI J0422+0047 & 2893541  & 65.715309  & +0.797050  & 4800 & 4.65 & 19.01, 19.23 & 1.982, 2.100 & projected quasar + galaxy \\ 
    DESI J0621+5227 & 1028529  & 95.478330  & +52.461052 & 1200 & 1.92 & 19.41, 19.82 & - & star + star \\
    DESI J0746+1344 & 10721197 & 116.648160 & +13.740173 & 1800 & 2.44 & 18.45, 20.04 & 3.105, 3.105 & lens        \\
    DESI J1556+7925 & 1006023  & 239.157775 & +79.427087 & 1500 & 0.93 & 18.75, 19.36 & 1.570 & likely lensed quasars        \\
    DESI J1759+3459 & 2197502  & 269.825027 & +34.991220 & 1800 & 2.30 & 19.56, 19.26 & 1.970, 1.982 & dual quasar \\
    DESI J1929+6009 & 10119139 & 292.255442 & +60.156748 & 2100 & 1.74 & 19.50, 20.24 & 3.277, 3.277 & dual quasar \\
    DESI J2121-0826 & 10912799 & 320.384762 & -8.436015  & 2400 & 1.34 & 19.63, 20.55 & 2.395 & lens        \\
    DESI J2303+0814 & 2587254  & 345.939046 & +8.248696  & 2100 & 1.82 & 20.71, 21.14 & 0.984, 0.928 & projected quasar \\ 
    \hline
    \end{tabular}
    
    \label{tab:obs}
\end{table*}

\begin{figure*}
\centering
    \includegraphics[width=1\textwidth]{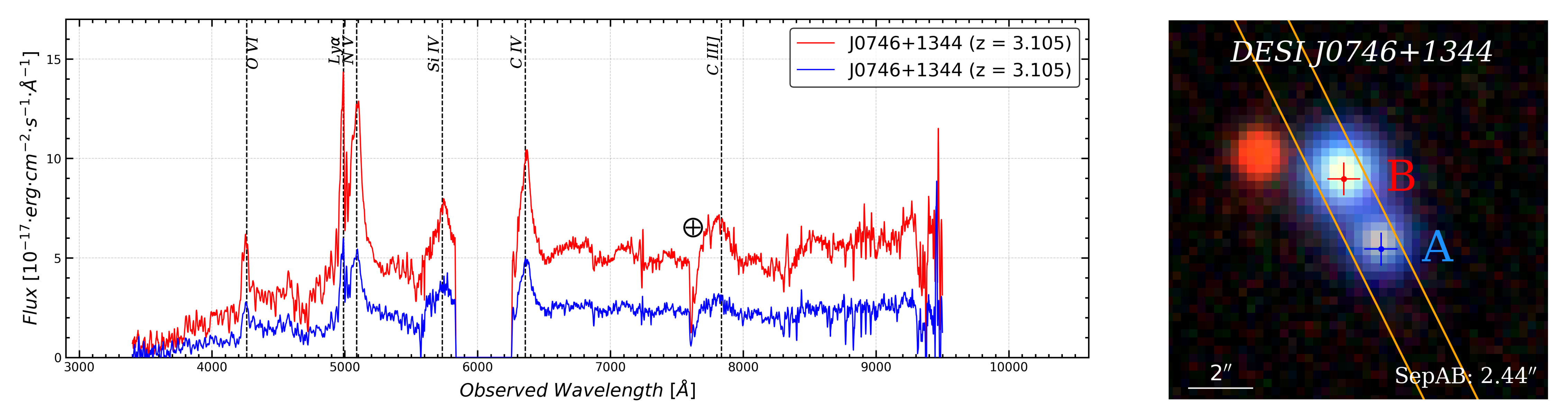}  
    \includegraphics[width=1\textwidth]{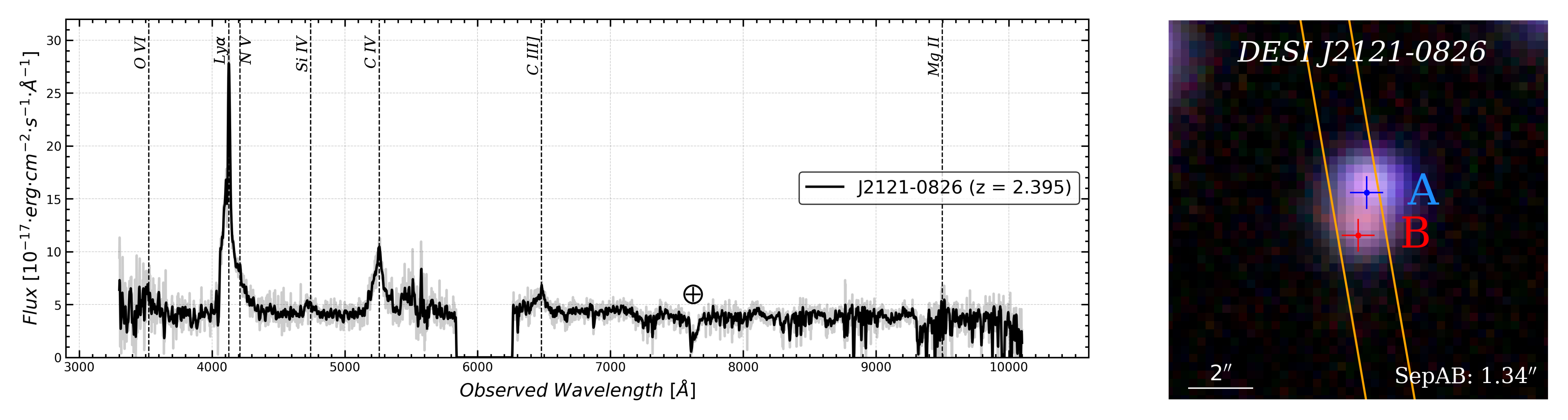} 
    \caption{The DBSP/P200 spectra and their corresponding cutout images of the two confirmed lensed quasars. The slit placement is labeled by the yellow lines.}
    \label{fig:p200-spec-lens}
\end{figure*}

\begin{figure*}
\centering
    \includegraphics[width=1\textwidth]{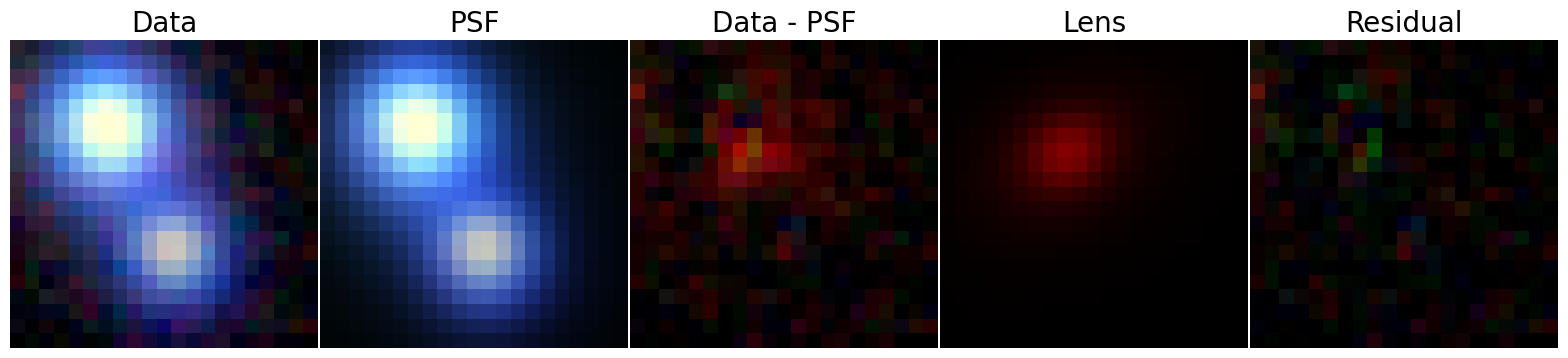}
    \includegraphics[width=1\textwidth]{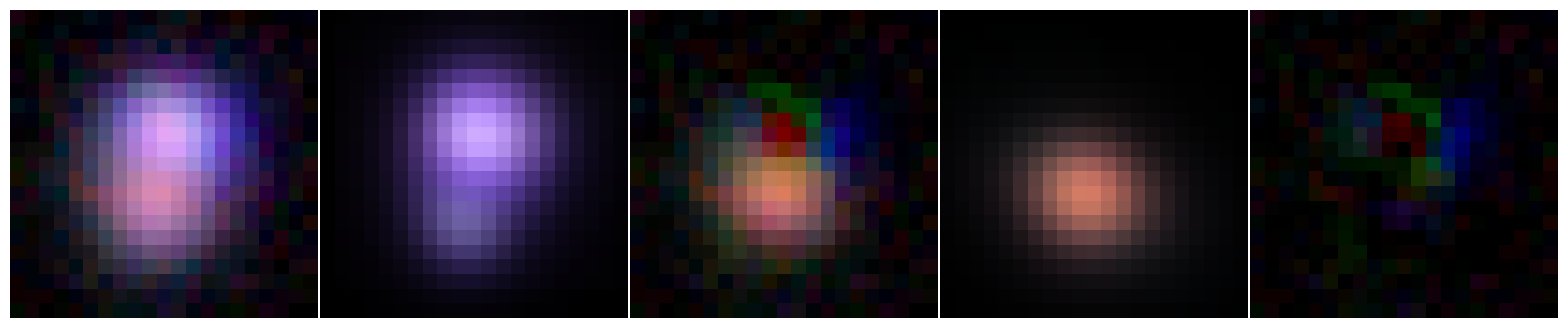}
    \caption{The image decomposition results in $grz-$band for our confirmed lensed quasars. First: J0746+1344, Second: J2121-0826. First column (A): original data. Second (B): light model for two images. Third (C): A-B; Fourth (D): light model for best-fitted lensing galaxy assuming Sérsic profile. Fifth (E): A-B-D.}
\label{fig:light-model}
\end{figure*}

\begin{table*}[]
\caption{Light modelling results of lensing galaxies in two confirmed lensed quasars. The meaning of the $n_s$, $r_s$, $q_s$, $\phi_s$ can be found at Appendix \ref{sec:lens_modelling}. $m_s$ is the modelled apparent magnitudes in different bands. The $+x$-axes is the $\phi_s=0$, while the counterclockwise direction is considered the positive direction.}
\centering
\begin{tabular}{c|c|ccccc}
\hline
name                        & band & $n_s$             & $r_s$             & $q_s$                             & $\phi_s$             & $m_s$              \\ \hline
\multirow{3}{*}{J0746+1344} & $g$  & \multicolumn{4}{c}{-}                                                                        & \textgreater{}23.0 \\
                            & $r$  & \multicolumn{4}{c}{-}                                                                        & \textgreater{}23.0 \\
                            & $z$  & $5.5\pm1.3$   & $0.9\pm0.2$   & $0.5\pm0.1$  & $27.0\pm11.0$  & $20.9\pm0.3$   \\
\multirow{3}{*}{J2121-0826} & $g$  & $5.9\pm0.2$   & $0.4\pm0.1$   & $0.7\pm0.1$  & $-20.9\pm5.4$  & $21.3\pm0.5$   \\
                            & $r$  & $3.2\pm0.2$   & $0.3\pm0.1$   & $0.6\pm0.1$  & $-26.0\pm2.4$  & $20.4\pm0.2$   \\
                            & $z$  & $5.0\pm0.7$   & $0.1\pm0.0$   & $0.7\pm0.1$  & $-21.1\pm2.2$  & $19.8\pm0.1$   \\ \hline
\end{tabular}
\label{tab:lens_gal}
\end{table*}

\begin{table}[]
\centering
\caption{The results of light modelling for multiple images of two confirmed lensed quasars are presented. Here, $\mu$ represents the magnification, while $m_{PSF}$ denotes the magnitude of the multiple images.}
\begin{tabular}{cccc}
\hline
name                          & $\mu$                & band & $m_{PSF}$                 \\ \hline
\multirow{3}{*}{J0746+1344 A} & \multirow{3}{*}{$2.150$} & $g$  & $20.41 \pm 0.13$  \\
                              &                      & $r$  & $20.07 \pm 0.11$  \\
                              &                      & $z$  & $19.99 \pm 0.08$  \\
\multirow{3}{*}{J0746+1344 B} & \multirow{3}{*}{$-1.610$} & $g$  & $19.26\pm 0.05$   \\
                              &                      & $r$  & $18.49 \pm 0.01$  \\
                              &                      & $z$  & $18.56 \pm 0.01$  \\
\multirow{3}{*}{J2121-0826 A} & \multirow{3}{*}{$2.868$} & $g$  & $20.01 \pm 0.07$  \\
                              &                      & $r$  & $19.92 \pm 0.07$  \\
                              &                      & $z$  & $19.30 \pm 0.06$ \\
\multirow{3}{*}{J2121-0826 B} & \multirow{3}{*}{$-1.678$} & $g$  & $21.68 \pm 0.56$  \\
                              &                      & $r$  & $21.50 \pm 0.49$  \\
                              &                      & $z$  & $21.07 \pm 0.45$  \\ \hline
\end{tabular}

\label{tab:psfs}
\end{table}

\begin{table*}[]
\caption{SIE parameters of two confirmed lensed quasars. The parameters are defined at Appendix \ref{sec:lens_modelling}.}
\centering
\begin{tabular}{cccccc}
\hline
name       & $\theta_E(\arcsec)$        & $q$               & $\phi(\deg)$              & RA($^\circ$)  & Dec($^\circ$)  \\ \hline
J0746+1344 & $1.208\pm 0.058$  & $0.622 \pm 0.102$ & $6.290\pm 3.865$    & 116.649    & 13.73955  \\
J2121-0826 & $0.770 \pm 0.028$ & $0.839 \pm 0.030$ & $-30.353 \pm 13.430$ & 320.384723 & -8.43616 \\ \hline
\end{tabular}

\label{tab:SIE}
\end{table*}

\begin{figure*}
    \centering
    \includegraphics[scale=0.6]{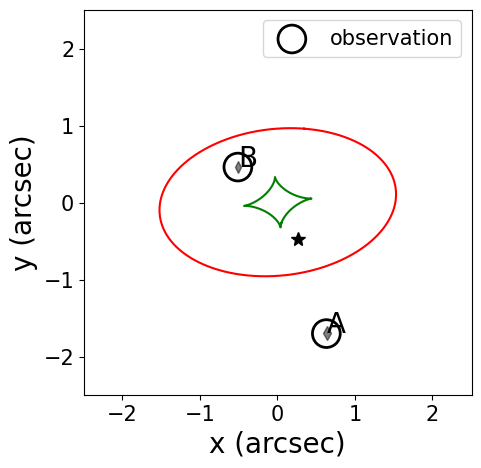}
    \includegraphics[scale=0.6]{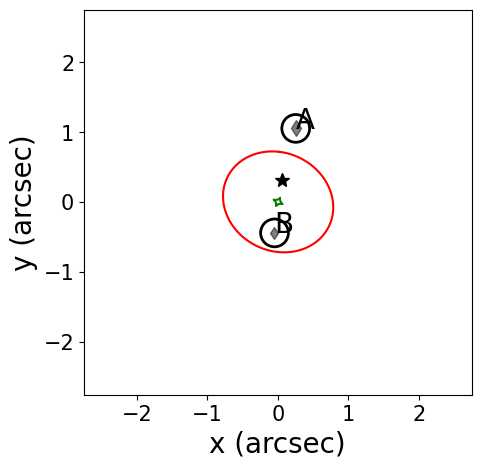}
    \caption{The modelled caustic and critical curves, along with the positions of images for the two confirmed lenses shown in Figure \ref{fig:p200-spec-lens}, are compared with the observed positions of multiple images. The plots are centred around the lens light centre, which is determined through light modelling and provided in Table \ref{tab:SIE}.}
    \label{fig:lens-model}
\end{figure*}

\begin{figure}
    \centering
    \includegraphics[scale=0.6]{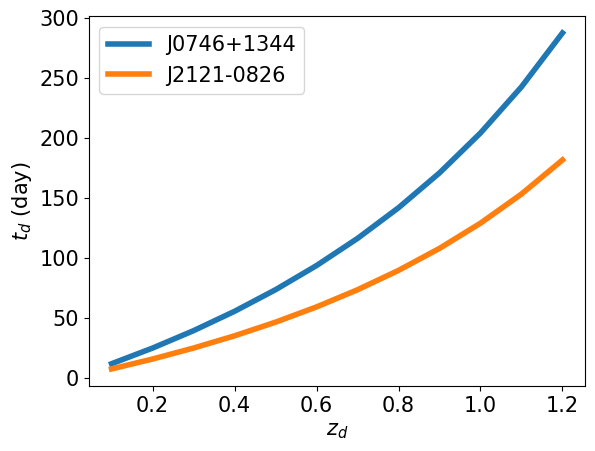}
    \caption{The time-delay as a function of $z_d$ of two confirmed lensed quasars.}
    \label{fig:td}
\end{figure}

\begin{figure*}
\centering
    \includegraphics[width=1\textwidth]{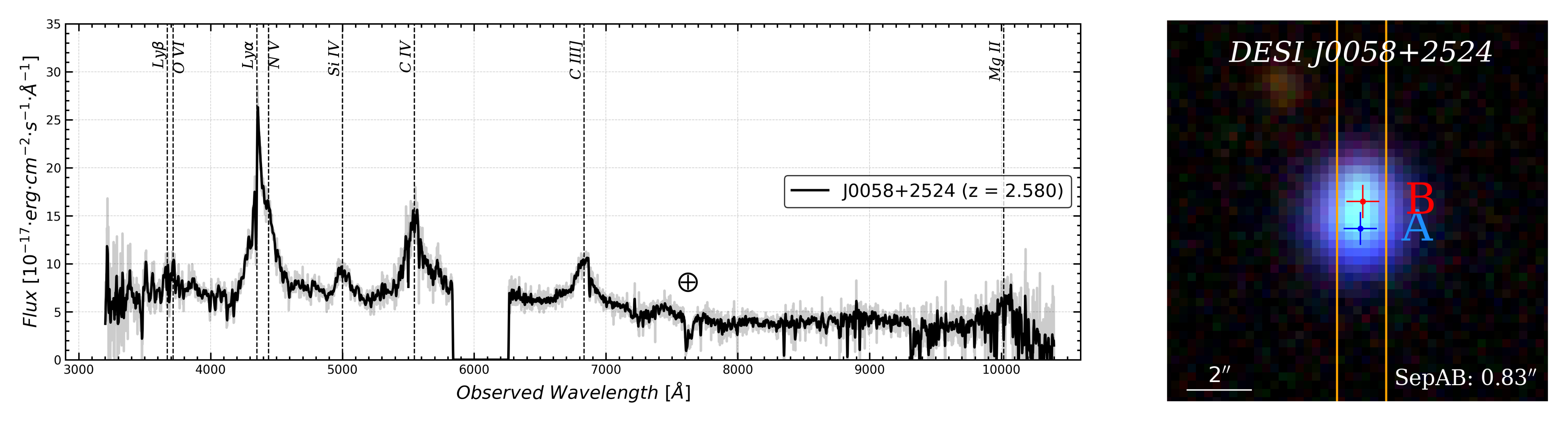}  
    \includegraphics[width=1\textwidth]{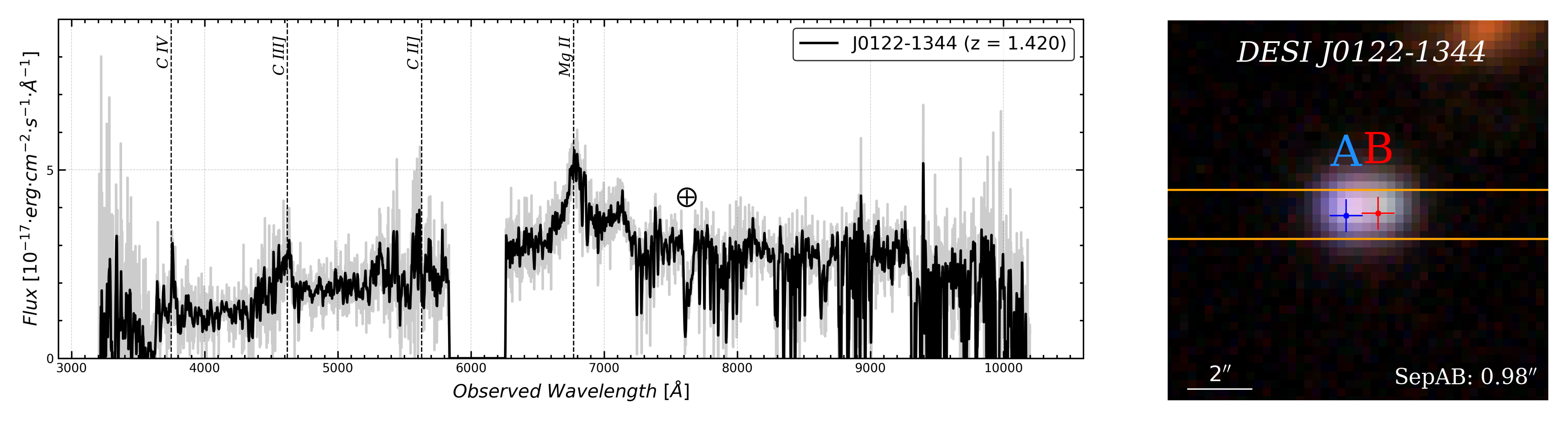}
    \includegraphics[width=1\textwidth]{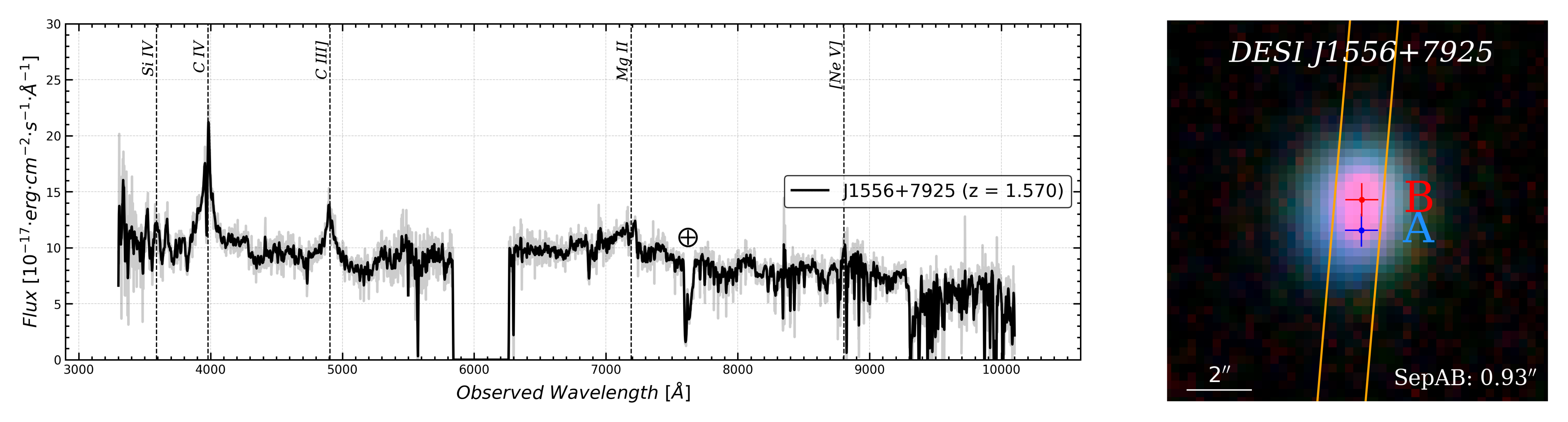}
    \caption{The DBSP/P200 spectra and their corresponding cutout images of the three likely lensed quasars. The slit placement is labeled by the yellow lines.}
    \label{fig:p200-spec-likelens}
\end{figure*}

\begin{figure}
    \centering
    \includegraphics[scale=0.4]{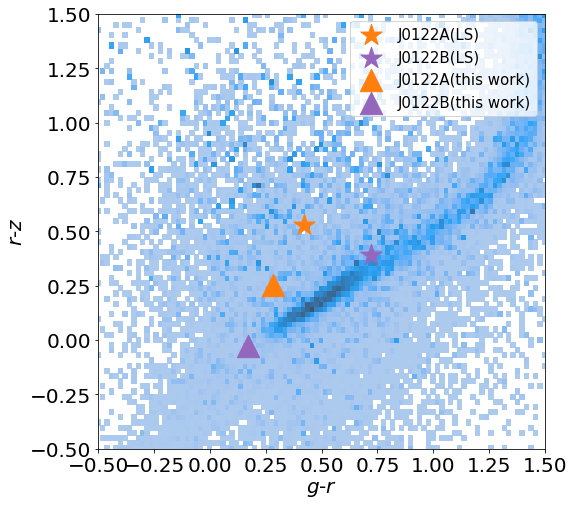}
    \caption{The colour-colour plot for J0122-1344 is displayed, with blue points indicating stars. The yellow and purple pentagrams represent J0122-1344A based on \LIS\ photometry. Furthermore, the yellow and purple triangles correspond to J0122-1344A and J0122-1344B, respectively, as determined by the light modelling results from this study, which utilises two PSFs along with one Sérsic profile (refer to Appendix \ref{sec:lens_modelling}).}
    \label{fig:J0122-star-locals}
\end{figure}

\begin{figure*}
\centering
    \includegraphics[width=1\textwidth]{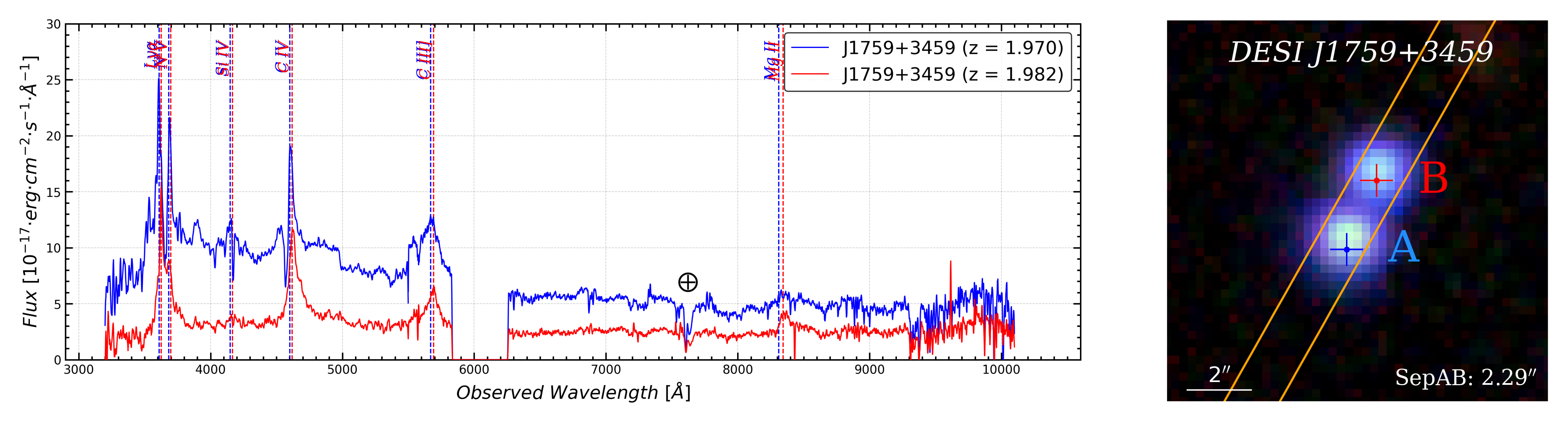}  
    \includegraphics[width=1\textwidth]{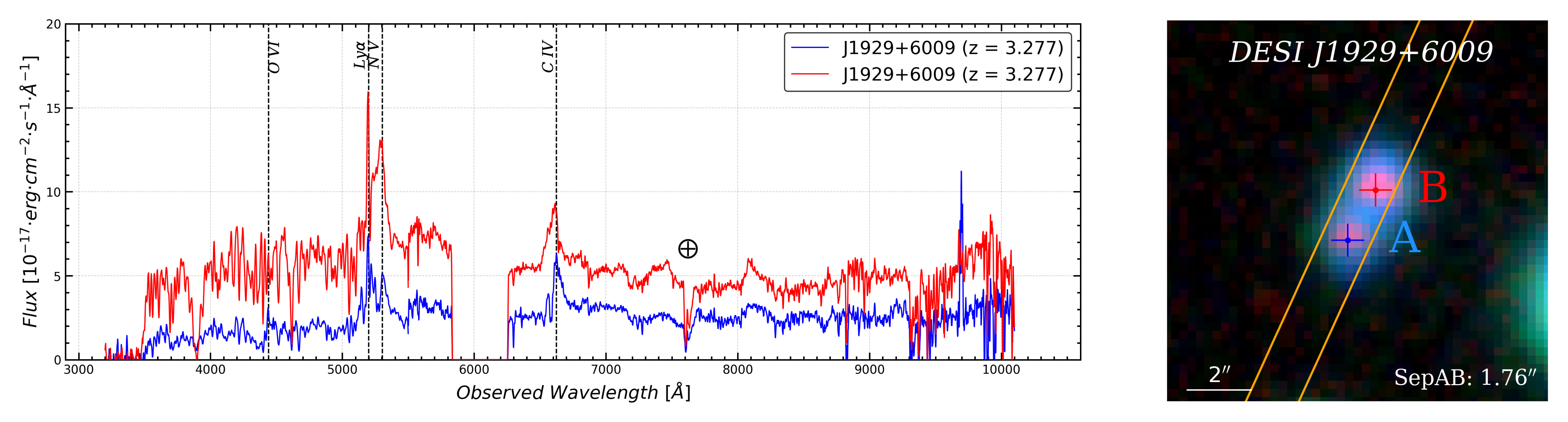} 
    \caption{The DBSP/P200 spectra and their corresponding cutout images of the two dual quasars. The slit placement is labeled by the yellow lines.}
    \label{fig:p200-spec-dq}
\end{figure*}

\begin{figure*}
\centering
    \includegraphics[width=1\textwidth]{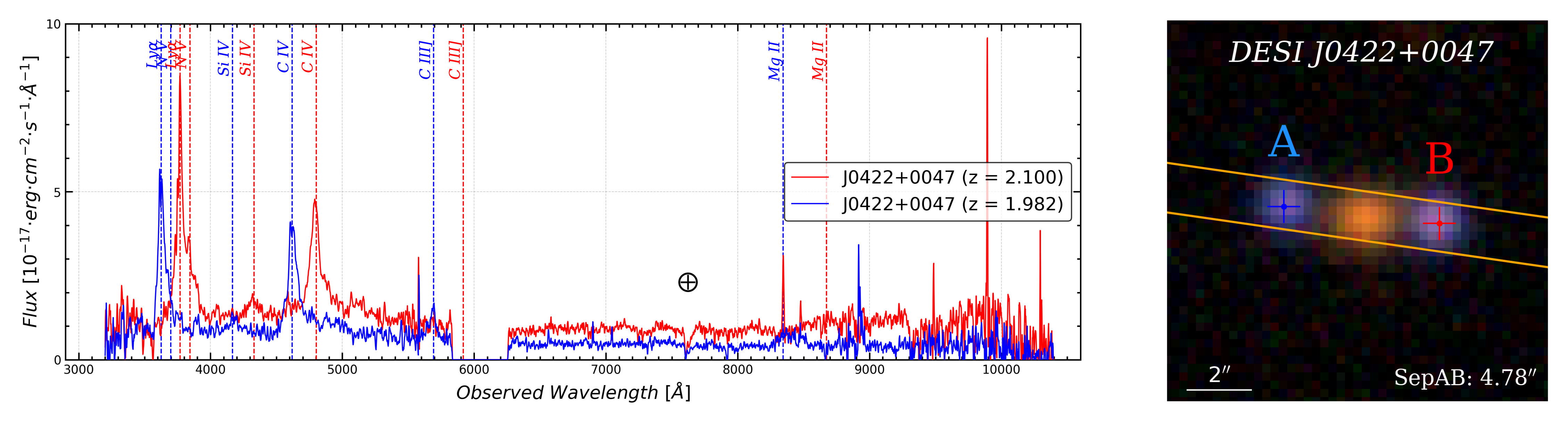}  
    \includegraphics[width=1\textwidth]{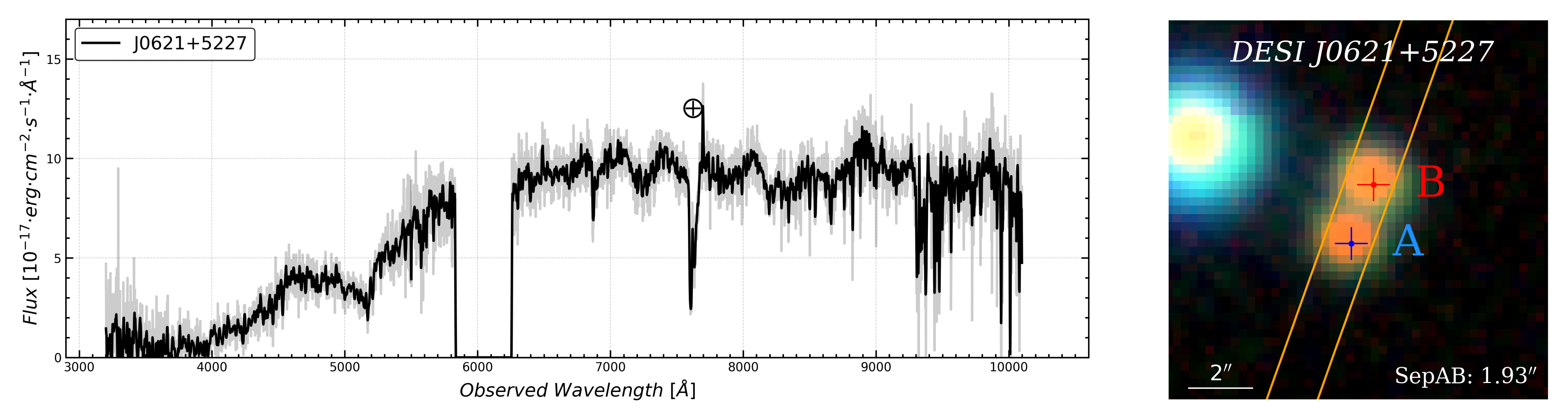} 
    \includegraphics[width=1\textwidth]{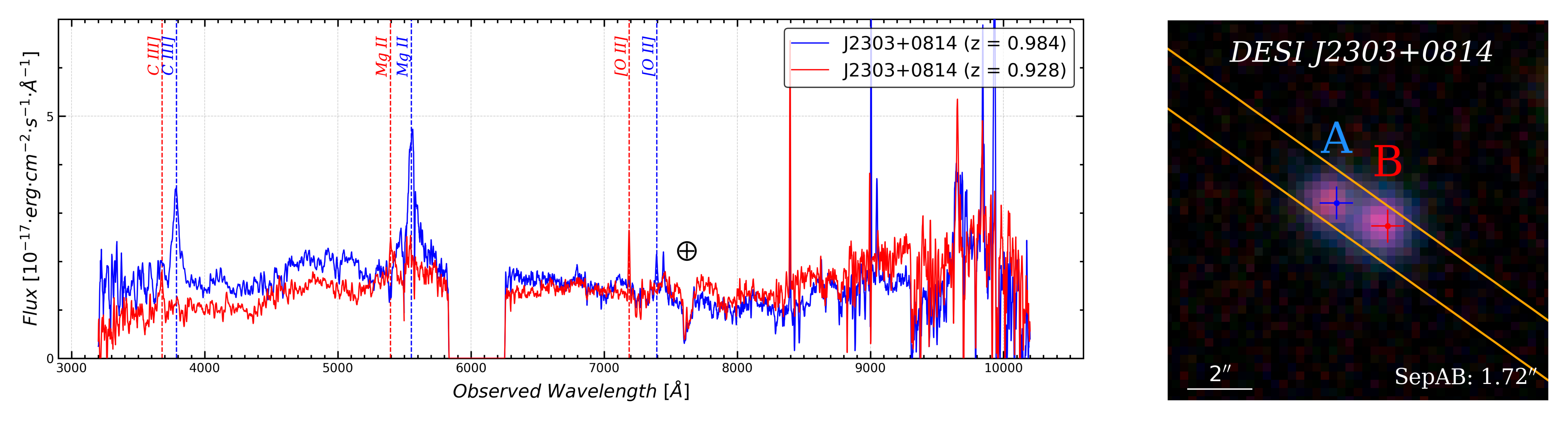} 
    \caption{The DBSP/P200 spectra and their corresponding cutout images of the two projected quasars and one star-star system (J0621+5227). The slit placement is labeled by the yellow lines. }
    \label{fig:p200-spec-pq+s}
\end{figure*}

\section{P200-DBSP Confirmation results}\label{sec:p200res}

The observation results of ten targets using P200 in Palomar Observatory are presented here. Combing spectra and \LIS, they were categorised into different classes, including lensed quasars (Section \ref{subsec:p200lq}), likely lensed quasars (Section \ref{subsec:p200likelq}), dual quasars (Section \ref{subsec:p200dq}), and projected quasars (Section \ref{subsec:p200pq}). Spectroscopic analysis reveals that J0621+5227 is a star-star system composed of K-type stars. As this stellar system is out of the scope of our primary scientific objectives, we will not discuss it further in the subsequent analysis. 

\subsection{Confirmed lensed quasars}\label{subsec:p200lq}

\subsubsection{J0746+1344}

Two point sources can be distinctly identified in the \LIS, with $r$-band magnitudes 18.45 and 20.04 respectively, separated by 2.$\arcsec$44. Spectroscopic analysis reveals two quasars at a redshift of $z$ = 3.105, exhibiting similar continuum and broad emission line profiles. This resemblance suggests that this is a gravitationally lensed system. In the light modelling procedure, the $grz$-band images are fitted well by two point spreading functions (PSFs) that represent the lensed quasar images, along with one Sérsic profile that represents the lensing galaxy. See top line of Figure \ref{fig:light-model} for the image reconstruction results, and Appendix \ref{sec:lens_modelling} for the details about light modelling.

Based on the spectral similarities between the two quasar images, the detection of an extended component consistent with a lensing galaxy, and the successful fitting of a two-PSF plus Sérsic model fit, we classify this system as a gravitational lens. The light distribution of lensing galaxy is fitted by a single Sérsic profile, the results are summarised at Table \ref{tab:lens_gal}. The results of multiple images are summarised at Table \ref{tab:psfs}.

Lens modelling is further performed using Singular Isothermal Ellipsoid \citep[SIE,][]{Kormann1994} model. We inputted the position of observed quasars, and assuming the SIE centre is same as light profile of lensing galaxy.  The result of SIE is summarised at Table \ref{tab:SIE}. Beside, the time-delay is estimated as a function of lens redshift, the result is given at Figure \ref{fig:td}. The magnification ($\mu$) of two images are also given in Table \ref{tab:psfs}. The detailed methodologies used in mass modelling, and time-delay estimation, can be found at Appendix \ref{sec:lens_modelling}.

\subsubsection{J2121-0826}

In \LIS, the $r$-band magnitudes of two images are 19.63 and 20.55 respectively, which are separated by 1.$\arcsec$34. Spectroscopic analysis of the DBSP data reveals a blended quasar spectrum at a redshift of $z$ = 2.395. The system exhibits two distinct detections in Gaia data, suggesting the presence of multiple images. In the residual of \LIS, a red component becomes visible after subtracting two PSFs. The $grz$-band composite image is well-modelled by simultaneously fitting a Sérsic profile positioned between the two PSFs, see the bottom line of Figure \ref{fig:light-model} for this, consistent with the presence of a lensing galaxy. Based on these evidence - the blended high-redshift quasar spectrum, multiple Gaia detections, and the presence of a red component consistent with a lensing galaxy - we classify this system as a gravitational lens.

The Sérsic parameters are presented in Table \ref{tab:lens_gal}. The PSF parameters can be found in Table \ref{tab:psfs}. The SIE parameters are detailed in Table \ref{tab:SIE}. Additionally, the time delay as a function of lens redshift is shown in Table \ref{fig:td}.

\subsection{Likely lensed quasars}\label{subsec:p200likelq}

\subsubsection{J0058+2524}

According to \LIS, the $r$-band magnitudes of two images are 18.94 and 19.71 respectively; two quasars are separated by 0.$\arcsec$83. We have exposed this for 1600s. Spectroscopic analysis of the DBSP data reveals a blended quasar spectrum at a redshift of $z$ = 2.580, accompanied by two distinct detections in Gaia data. When modelled using a combination of two PSFs and one Sérsic profile, a potential lensing galaxy is detected between the two PSFs, albeit at a low significance level of $\sim 1\sigma$.

The presence of a blended quasar spectrum at \z{2.58} and multiple Gaia detections are promising indicators of a gravitational lensing system. However, the low detection significance of the lensing galaxy introduces uncertainty into this classification. Given the ambiguity regarding the lensing galaxy detection, we conclude that deeper, higher-resolution imaging is necessary to conclusively confirm or refute this gravitational lens candidate.

\subsubsection{J0122-1344}

\begin{table*}[]
\centering
\caption{Magnitudes of J0122-1344, including both the Sérsic and PSF components, are presented. This includes photometry from this study as well as data from \LIS. The LS recognised this system as only two PSFs, the Sérsic component is not suitable.}
\begin{tabular}{ccccccc}
\hline
               & $g$(LS) & $r$(LS) & $z$(LS) & $g$(this work)   & $r$(this work) & $z$(this work) \\ \hline
Sérsic  & \multicolumn{3}{c}{-}       & \textgreater{}23 & 20.51          & 19.48          \\
J0122A         & 20.32   & 19.9    & 19.37   & 20.29            & 20.04          & 19.78          \\
J0122B         & 21.04   & 20.32   & 19.93   & 21.08            & 20.87          & 20.90          \\ \hline
\end{tabular}

\label{tab:J0122-light}
\end{table*}

From \LIS, two point-like components with different colours are visible. In the residual imaging of \LIS, an extended yellow component is  clearly seen. We performed light modelling on this system, and the results are summarised in Table \ref{tab:J0122-light}. Here are three key observational facts: Firstly, typical quasar broad-line emissions, such as $C_{IV}$ and $Mg_{II}$, can be identified from the blended spectrum, indicating the presence of at least one quasar. However, the continuum is very different from typical AGN continuum. Secondly, an extended galaxy is clearly visible from Sérsic + two PSF light modelling, with results indicating the Sérsic galaxy with magnitudes of $m_r = 19.48$ and $m_z = 20.51$. Thirdly, Figure \ref{fig:J0122-star-locals} presents the $g$-$r$ vs $r$-$z$ plot for two images. Both the magnitudes from DESI-LS and our modelling are shown. The results from DESI-LS suggest the image B is more possible a star, image A is more possible a quasars, while our light modelling indicates that both images are quasars.

These observations can be explained by two scenarios.
\begin{itemize}
    \item Lensing Scenario. Images A and B are doubly-imaged quasars. A bright lensing galaxy significantly contributes to the spectrum, causing the continuum to deviate from a typical quasar spectrum. The color difference between the two images can be attributed to the blended light from the lensing galaxy.
    \item Quasar-Galaxy-Star Scenario. The two images represent a quasar-star pair, with a galaxy coincidentally positioned between them. According to Figure \ref{fig:J0122-star-locals}, the lensing scenario is more plausible because the colour of two images are differ from typical star locations.
\end{itemize}

To distinguish between these scenarios, improved spectroscopic observations that can capture two distinct spectra from the two point sources are necessary.

\subsubsection{J1556+7925}

The DBSP data reveals a blended quasar spectrum at $z$=1.507, and two distinct Gaia detections support the hypothesis that this system is a lensed quasar. However, no lensing galaxy is detectable after subtracting two PSFs in the Legacy Survey imaging. We believe deeper imaging is necessary to confirm this candidate.

\subsection{Dual quasars}\label{subsec:p200dq}

\subsubsection{J1759+3459}

Spectroscopic analysis reveals the presence of two distinct quasars at redshifts $z = 1.970$ and $z = 1.982$, respectively. The difference in redshifts effectively rules out the possibility of gravitational lensing. We classify this system as a binary quasar pair with an angular separation of $\Delta \theta = 2.3\arcsec$, corresponding to a projected physical separation of 19.57 kpc at $z = 1.970$. The line-of-sight velocity difference between the two quasars is $\Delta v = 1208\pm 14 \ km/s$.

\subsubsection{J1929+6009}

Spectroscopic analysis reveals the presence of two distinct quasars at a redshift of $z = 3.277$, separated by an angular distance of $\Delta \theta = 1.34\arcsec$. The possibility of gravitational lensing can be confidently ruled out due to the significant differences observed in the continuum spectra of the two quasars. These spectral variations may be attributed to differences in the accretion disk properties, dust extinction, or host galaxy contributions between the two quasars. We classify this system as a dual quasar pair with a projected physical separation of 13.4 kpc at the observed redshift, featuring a velocity difference of less than $10\ km/s$.

\subsection{Projected quasars}\label{subsec:p200pq}

\subsubsection{J0422+0047}

Initially thought to be a gravitationally lensed quasar system with the central yellow component acting as the lensing galaxy, this intriguing configuration has been revealed to be more complex. Spectroscopy of the two quasar objects and the central component has definitively ruled out the lensing scenario due to the significant quasar redshift differences observed. 

The large $\Delta v$ of $11633 \pm 14 \ km/s$ between the quasars leads us to classify this system as a chance alignment of two projected quasars with an intervening galaxy. Despite not being a gravitational lens, this serendipitous alignment enables us to investigate the CGM properties of both the quasar host galaxies and the intervening galaxy. 

\subsubsection{J2303+0814}

Spectroscopic analysis reveals two distinct quasars with significantly different redshifts of $z = 0.948$ and $z = 0.928$, respectively. The substantial difference in their continuum spectra, coupled with the redshift difference, definitively rules out the possibility of gravitational lensing. The redshift difference between the two quasars is $\Delta z = 0.020$, which translates to a line-of-sight velocity difference of $\Delta v = 8583 \pm 22\ km/s$. This large velocity separation far exceeds what would be expected for physically associated quasars, further supporting our classification.

\section{Inspection on available data-sets}\label{sec:specdataset}

We conducted a cross-match between H22 systems and several spectroscopic datasets with radius equals 1$\arcsec$ (DESI-EDR and SDSS-DR16). Our analysis revealed new dual quasars (Section \ref{subsec:DQ_in_dataset}) and projected quasars (Section \ref{subsec:PQ_in_dataset}). To the best of our knowledge, these systems have not been previously discussed in the literature.

\begin{figure*}
\centering

    \includegraphics[scale=0.45]{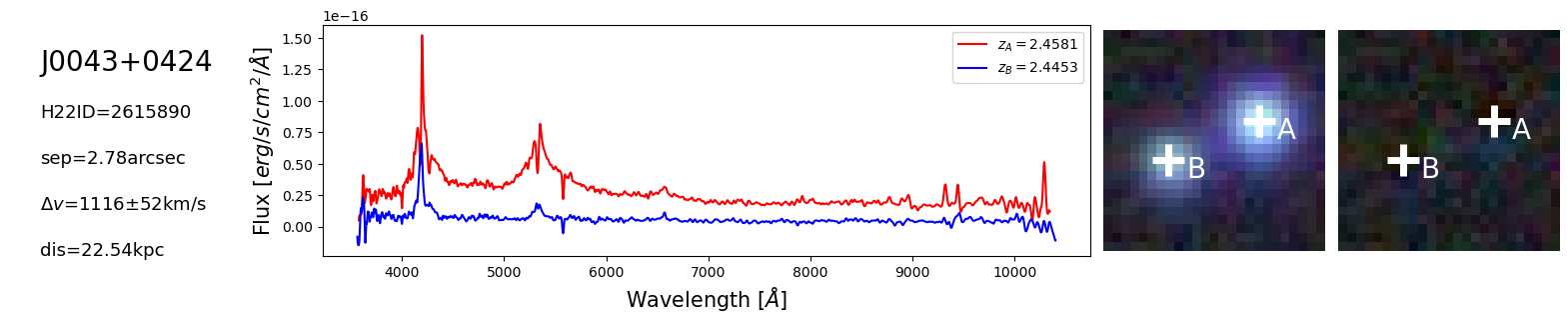}
    \includegraphics[scale=0.45]{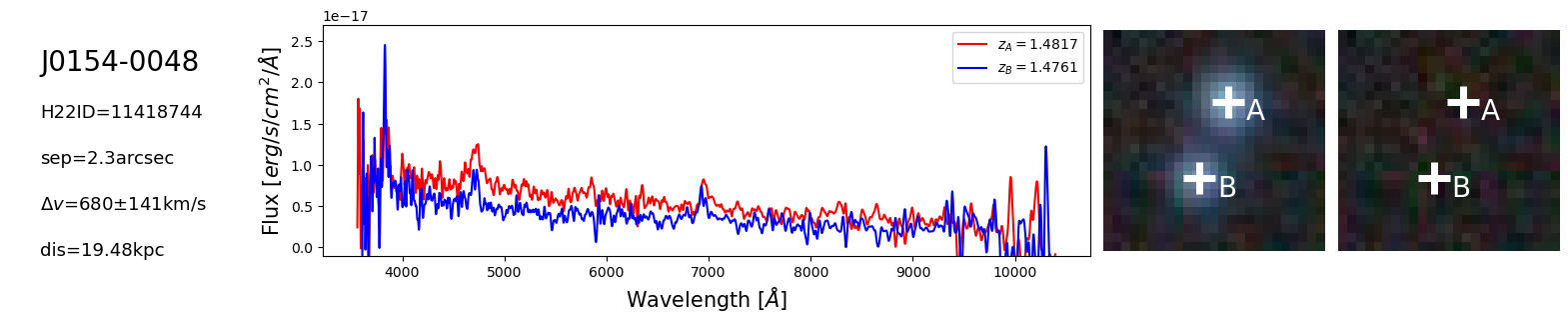}
    \includegraphics[scale=0.45]{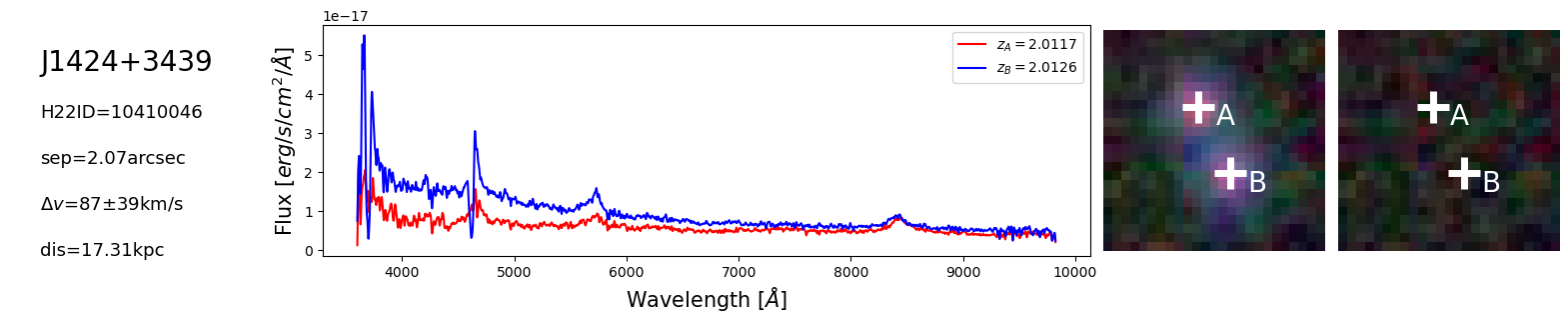}
    \includegraphics[scale=0.45]{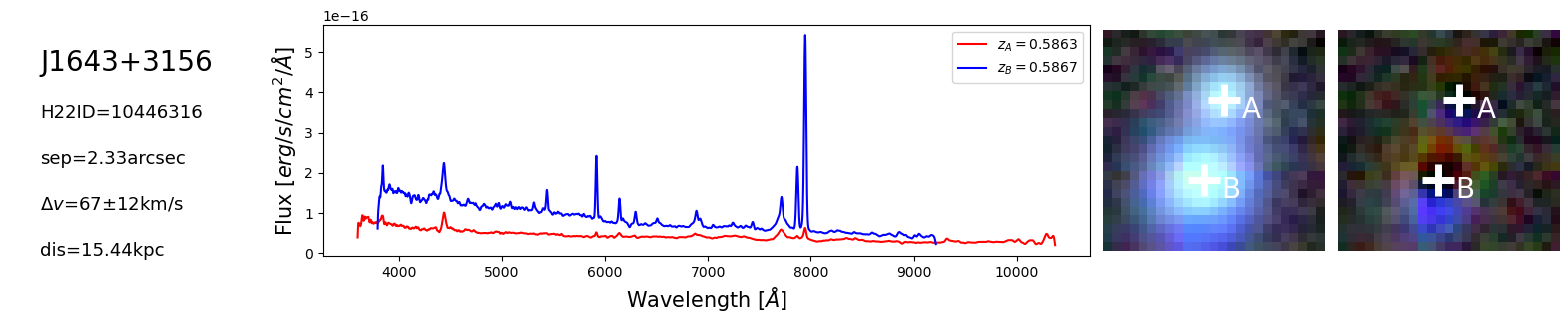}

    \caption{We present data on four dual quasars, verified using publicly available spectral datasets.
On the left, we provide essential details: $\Delta v$ represents the velocity difference, `sep' denotes the image separation of the two quasars, and `dis' indicates the projected separation in the line-of-sight direction at lower redshift. The second column displays the spectra, where the fluxes have been smoothed using a Gaussian kernel with a standard deviation of 5 $\AA$. The third column features the $grz$ color image from the DESI-LS DR10, while the final column shows the residual image derived from DESI-LS DR10.}
    \label{fig:specs_dataset_dpQ}
\end{figure*}

Table \ref{tab:para_summary_dataset_dpQ} summaries the key information for these quasar systems, including their Right Ascension (RA), Declination (Dec), data source, angular separation, redshifts, and associated redshift errors. Figure \ref{fig:specs_dataset_dpQ} presents the spectra of the dual quasars, along with their velocity differences ($\Delta_v$) and projected separations.

\subsection{Dual quasars}\label{subsec:DQ_in_dataset}

\subsubsection{J0043+0424}

Spectroscopic analysis from BOSS reveals that two of the systems exhibit characteristic broad emission lines typical of AGN. Notably, a self-absorption feature is observed in the C$_{IV}$ emission line of quasar A, while it is absent in quasar B. This distinct spectral difference strongly suggests that we are observing a dual quasar system at \z{2.458} and \z{2.445} separated by \T{2.78} rather than a gravitationally lensed quasar. On the other hand, the small $\Delta v = 1116 \pm 25 \ km/s$ suggest they are not projected quasars. Thus we classify it as dual quasars.

\subsubsection{J0154-0048}

The spectra from BOSS exhibit typical AGN features, including prominent broad emission lines such as C$_{\text{III]}}$, with slightly different redshifts between the two quasars.   The redshift difference corresponding to a $\Delta v$ of $680\pm 141\ km/s$. According to studies by \cite{Oguri2010} and \cite{Cao2023}, the $z$-band magnitude ($m_z$) for such lensing systems typically ranges from 14 to 21, which is brighter than the detection limit of the Legacy Imaging Surveys. Given this, we would expect to detect a deflector if this were a gravitational lens system. However, no such deflector is observed. Consequently, we classify this system as a dual quasar with a separation of \T{2.3} or \d{19.48} at \z{1.4817}.

\subsubsection{J1424+3439}

Spectroscopic analysis from DESI reveals two sources exhibiting quasar characteristics at slightly different redshifts,  which can be transferred to a $\Delta v$ of $87 \pm 39 \ km/s$. Furthermore, no evidence of a gravitational lens deflector is detected in the residual images from the Legacy Survey. Consequently, we classify this system as a dual quasar pair with a projected separation of \T{2.07} or \d{17.31} at \z{1.4716}. 

\subsubsection{J1643+3156}

Spectroscopic analysis from SDSS and BOSS confirms that components A and B are both quasars at approximately \z{0.586}. According to studies by \cite{Oguri2010} and \cite{Cao2023}, the $m_z$ of a potential lensing galaxy for a system with $z_s=$0.586 should be brighter than 17. Such a galaxy would be unambiguously detectable in the Legacy Imaging Surveys. Given the absence of any visible deflector galaxy in the LIS data, we can rule out the lensing scenario for this system. Therefore, we classify this system as a dual quasar pair with a projected separation of \T{2.33} or \d{15.44} at \z{0.5853}.  The velocity difference is $67 \pm 12\ km/s$.

\subsection{Projected quasars}\label{subsec:PQ_in_dataset}

\subsubsection{J0134+3308}

The spectra under examination are from the BOSS. This particular case presents an intriguing projected quasar pair with an angular separation of 3.56$\arcsec$. While the two quasars exhibit remarkably similar colours, their redshifts differ significantly, with one at \z{0.7064} and the other at \z{1.1359}. Interestingly, both quasars show evidence of their host galaxies in the spectral residuals after subtracting the quasar contribution. 

\subsubsection{J0843+4733}

The broad-line emissions like C$_{IV}$, C$_{III}$, Mg$_{II}$ are identified for both quasars. Of particular interest is the spectrum of quasar A, which displays a significant C$_{III}$ absorption feature at $\sim 4000$Å. This absorption is potentially  due to the presence of quasar B at a lower redshift. Such an configuration presents a valuable opportunity for investigating the CGM of the quasar host galaxy.

\subsubsection{J1201-0117}

The DESI-EDR spectra reveal two quasars at markedly different redshifts: one at \z{2.5797} (quasar A) and another at \z{1.0653} (quasar B). A emission of 9400 Å seems not a AGN typical emission, probably from atmosphere.

\subsubsection{J1246+5030}

The SDSS spectra reveal two quasars at significantly different redshifts: one at \z{2.7307} (quasar A) and another at \z{2.1129} (quasar B). This substantial redshift difference creates an intriguing cosmic alignment that offers a unique opportunity for studying quasar absorption systems. In the spectrum of the higher-redshift quasar A, a  significant C$_{IV}$ absorption feature is  clearly seen. This absorption is particularly interesting as it may be  due to the presence of quasar B, which lies at a lower redshift along the same line of sight.

\subsubsection{J1351+5224}

BOSS spectroscopic observations reveal two quasars with significantly different redshifts (\z{3.2027} and \z{0.9747}). The spectrum of quasar A exhibits a prominent, broad Lyman-$\alpha$ emission line accompanied by the characteristic Lyman-$\alpha$ forest.

\subsubsection{J1537+3649}

BOSS spectroscopic observations reveal two quasars with distinctly different redshifts (\z{1.3850} and \z{1.0629}). Based on this significant redshift discrepancy, we classify this system as a projected quasar pair separated by \T{2.53}.

\subsubsection{J1603+5449}

DESI-EDR spectroscopic observations reveal two quasars with distinctly different redshifts (\z{1.5171} and \z{0.3451}). Based on this significant redshift discrepancy, we classify this system as a projected quasar pair separated by \T{2.24}.

\subsubsection{J1655+3408}

Spectroscopic analysis from the DESI-EDR confirms that quasars A and B are distinct objects. They are separated by or a projected separation of \d{28.25} at \z{1.7316}. The individual redshifts of the quasars (\z{1.7316} and \z{1.6931} respectively) yield a velocity difference of $4261\pm 37\ km/s$, which exceeds conventional boundary for dual quasars \citep[$2,000\ km/s $,][]{Hennawi2010}. Consequently, we classify this  system as projected quasar pair.

\subsubsection{J1758+6654}

DESI-EDR spectroscopic data reveals a pair of quasars with markedly different redshifts (\z{1.9550} and \z{2.9151}). The spectrum of the higher-redshift quasar (B) exhibits a significant Mg$_{II}$ absorption feature at $\sim 8400$Å. Intriguingly, this absorption may be attributed to the presence of the lower-redshift quasar (A), suggesting the light emitted by quasar B is absorbed by A, despite their considerable redshift difference.

\begin{table*}[h]
\caption{The information of the confirmed dual quasars (DQ) and projected quasars (PQ) from publicly available data-sets.}
\tiny
\centering
\begin{tabular}{ccccccccc}
\hline
Name        & H22-ID                    & RA($^\circ$)        & Dec($^\circ$)         & magr    & Separation ($^{\prime\prime}$) & Quasar Redshift (z)        & Dataset   & outcome             \\ \hline
J0043+0424A & \multirow{2}{*}{2615890}  & 10.970369   & 4.4076394   & 20.0954 & \multirow{2}{*}{2.78}          & $2.4581 \pm 0.0003$        & BOSS-DR16 & \multirow{2}{*}{DQ} \\
J0043+0424B &                           & 10.969656   & 4.4073359   & 21.1319 &                                & $2.4453 \pm 0.0005$        & BOSS-DR16 &                     \\ \hline
J0154-0048A & \multirow{2}{*}{11418744} & 28.709636   & -0.80441733 & 21.4140 & \multirow{2}{*}{2.30}          & $1.4817 \pm 0.0004$        & BOSS-DR16 & \multirow{2}{*}{DQ} \\
J0154-0048B &                           & 28.70941    & -0.80501563 & 21.9943 &                                & $1.4761 \pm 0.0011$        & BOSS-DR16 &                     \\ \hline
J1424+3439A & \multirow{2}{*}{10410046} & 216.0092993 & 34.65426983 & 21.6626 & \multirow{2}{*}{2.07}          & $2.0117 \pm 0.0003$        & DESI-EDR  & \multirow{2}{*}{DQ} \\
J1424+3439B &                           & 216.009599  & 34.65375044 & 21.6028 &                                & $2.0126 \pm 0.0002$        & DESI-EDR  &                     \\ \hline
J1643+3156A & \multirow{2}{*}{10446316} & 250.79745   & 31.939072   & 19.5198 & \multirow{2}{*}{2.33}          & $0.5863 \pm 0.0001$        & BOSS-DR16 & \multirow{2}{*}{DQ} \\
J1643+3156B &                           & 250.79726   & 31.938444   & 18.1591 &                                & $0.5867 \pm 0.00003$       & SDSS-DR16 &                     \\ \hline
J0134+3308A & \multirow{2}{*}{10428646} & 23.570643   & 33.1485     & 19.1408 & \multirow{2}{*}{3.56}          & $1.1359 \pm 0.0006$        & BOSS-DR16 & \multirow{2}{*}{PQ} \\
J0134+3308B &                           & 23.569548   & 33.148871   & 19.9502 &                                & $0.7064 \pm 0.0001$        & BOSS-DR16 &                     \\ \hline
J0843+4733A & \multirow{2}{*}{10232318} & 130.74015   & 47.56241    & 20.0026 & \multirow{2}{*}{3.42}          & $1.6778 \pm 0.0002$        & BOSS-DR16 & \multirow{2}{*}{PQ} \\
J0843+4733B &                           & 130.73904   & 47.561825   & 19.4983 &                                & $1.5549 \pm 0.0003$        & SDSS-DR16 &                     \\ \hline
J1201-0117A & \multirow{2}{*}{11421719} & 180.1501026 & -1.29273836 & 20.9270 & \multirow{2}{*}{2.26}          & $2.5797 \pm 0.0002$        & DESI-EDR  & \multirow{2}{*}{PQ} \\
J1201-0117B &                           & 180.1494852 & -1.29285373 & 20.7197 &                                & $1.0653 \pm 0.0002$        & DESI-EDR  &                     \\ \hline
J1246+5030A & \multirow{2}{*}{2111676}  & 191.55732   & 50.51334    & 19.4188 & \multirow{2}{*}{2.61}          & $2.7307 \pm 0.0005$        & BOSS-DR16 & \multirow{2}{*}{PQ} \\
J1246+5030B &                           & 191.55674   & 50.513965   & 19.3434 &                                & $2.1129 \pm 0.0002$        & BOSS-DR16 &                     \\ \hline
J1351+5224A & \multirow{2}{*}{1028812}  & 207.62799   & 52.401211   & 19.6030 & \multirow{2}{*}{2.66}          & $3.2027 \pm 0.0005$        & BOSS-DR16 & \multirow{2}{*}{PQ} \\
J1351+5224B &                           & 207.62737   & 52.401845   & 20.7675 &                                & $0.9747 \pm 0.0002$        & BOSS-DR16 &                     \\ \hline
J1537+3649A & \multirow{2}{*}{10343393} & 234.13051   & 36.819902   & 20.7280 & \multirow{2}{*}{2.53}          & $1.3850 \pm 0.0005$        & BOSS-DR16 & \multirow{2}{*}{PQ} \\
J1537+3649B &                           & 234.13123   & 36.819501   & 21.2257 &                                & $1.0629 \pm 0.0006$        & BOSS-DR16 &                     \\ \hline
J1603+5449A & \multirow{2}{*}{10143011} & 240.8114993 & 54.81770645 & 21.5314 & \multirow{2}{*}{2.24}          & $1.5171 \pm 0.0003$        & DESI-EDR  & \multirow{2}{*}{PQ} \\
J1603+5449B &                           & 240.8120508 & 54.81824093 & 21.9013 &                                & $0.3451 \pm 0.0002$        & DESI-EDR  &                     \\ \hline
J1655+3408A & \multirow{2}{*}{10414988} & 253.7660233 & 34.13442421 & 21.2080 & \multirow{2}{*}{3.34}          & $1.7316 \pm 0.0001$        & DESI-EDR  & \multirow{2}{*}{PQ} \\
J1655+3408B &                           & 253.7655554 & 34.13358145 & 20.2031 &                                & $1.6931 \pm 0.0003$        & DESI-EDR  &                     \\ \hline
J1758+6654A & \multirow{2}{*}{10040521} & 269.4174023 & 66.90915019 & 21.4507 & \multirow{2}{*}{2.40}          & $1.9550 \pm 0.0002$        & DESI-EDR  & \multirow{2}{*}{PQ} \\
J1758+6654B &                           & 269.4186132 & 66.90868133 & 22.2050 &                                & $2.9151 \pm 0.0003$        & DESI-EDR  &                     \\ \hline
\end{tabular}
\label{tab:para_summary_dataset_dpQ}
\end{table*}

\begin{figure*}
\centering
    \includegraphics[scale=0.45]{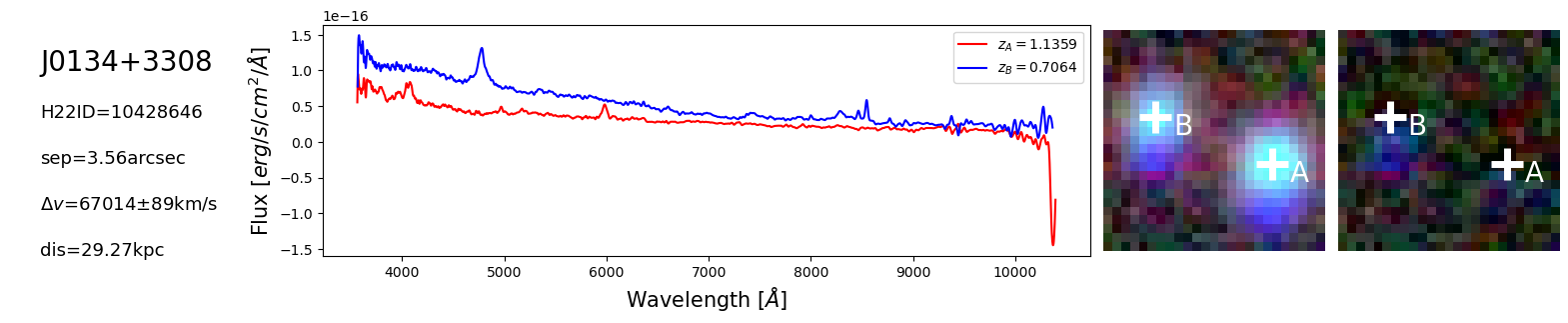}
    \includegraphics[scale=0.45]{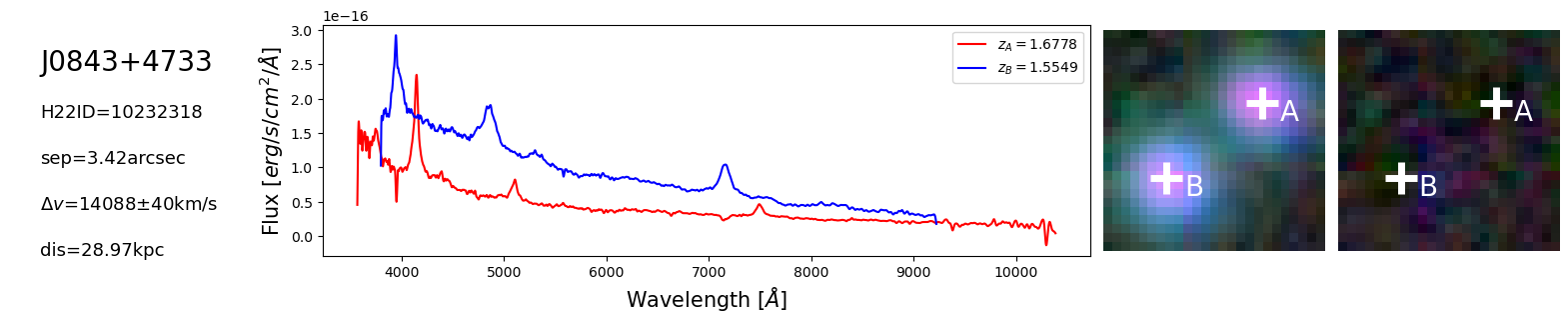}
    \includegraphics[scale=0.45]{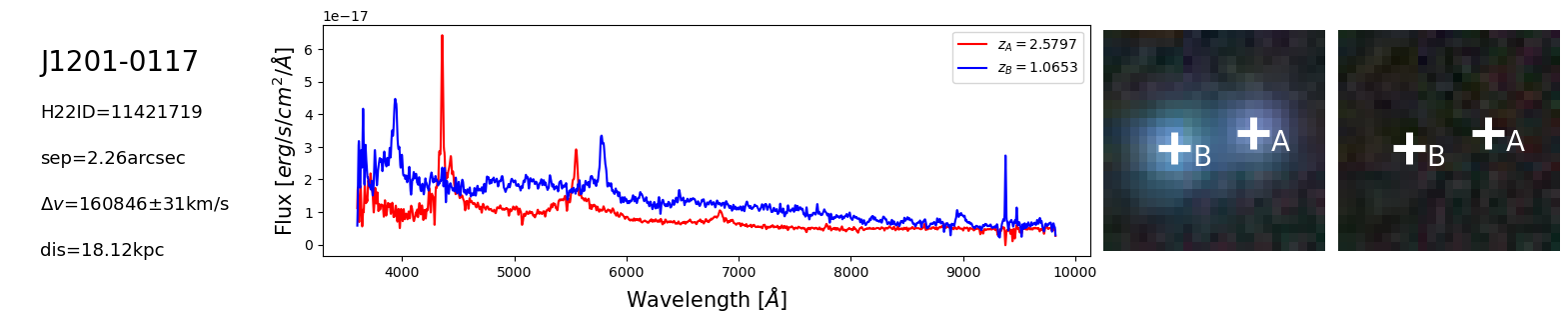}
    \includegraphics[scale=0.45]{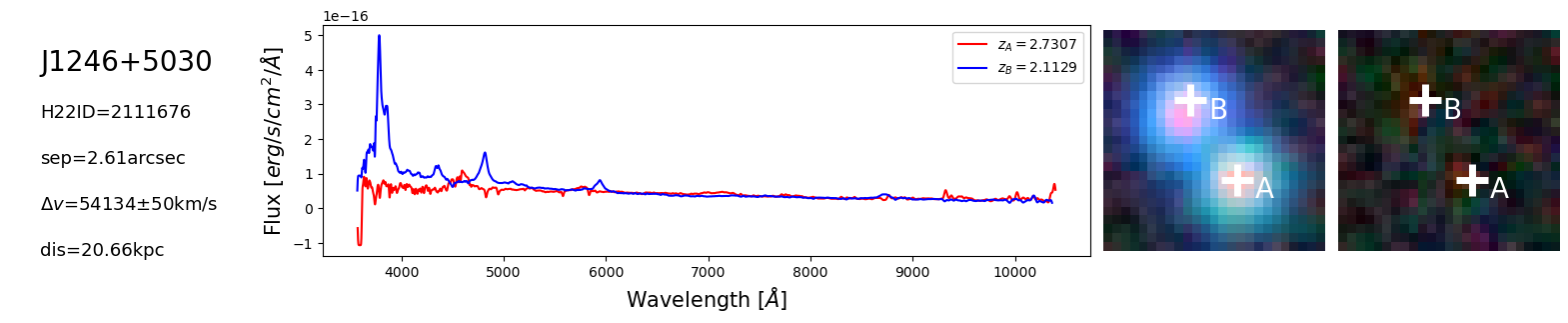}
    \includegraphics[scale=0.45]{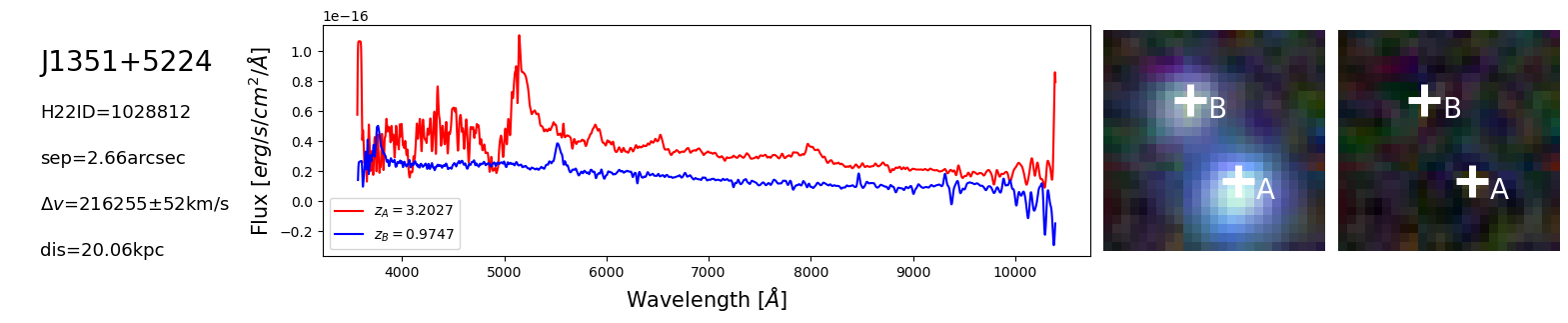}
    \includegraphics[scale=0.45]{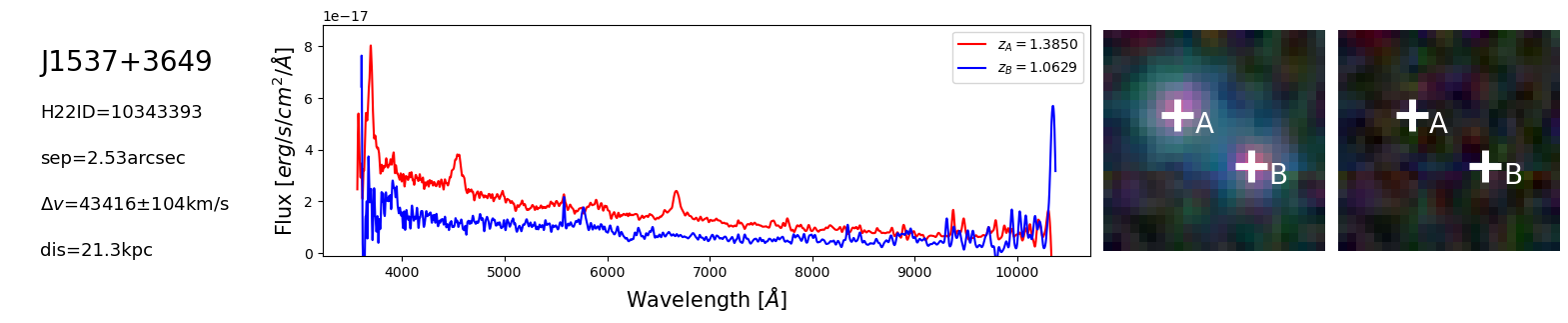}
    \caption{We present data on nine projected quasars, verified using publicly available spectral datasets. The configuration of this plot is the same with Figure \ref{fig:specs_dataset_dpQ}.}
\end{figure*}

\addtocounter{figure}{-1}

\begin{figure*}
\centering
    \includegraphics[scale=0.45]{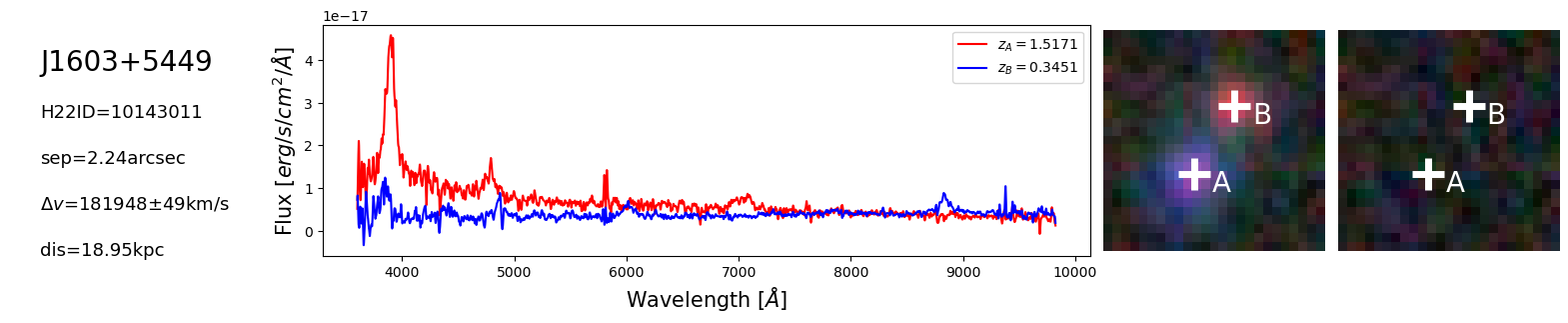}
    \includegraphics[scale=0.45]{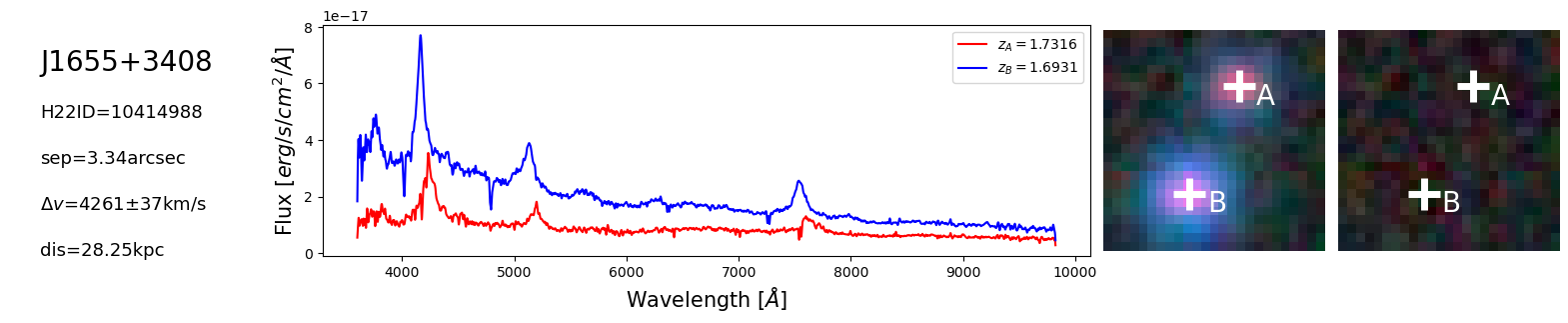}
    \includegraphics[scale=0.45]{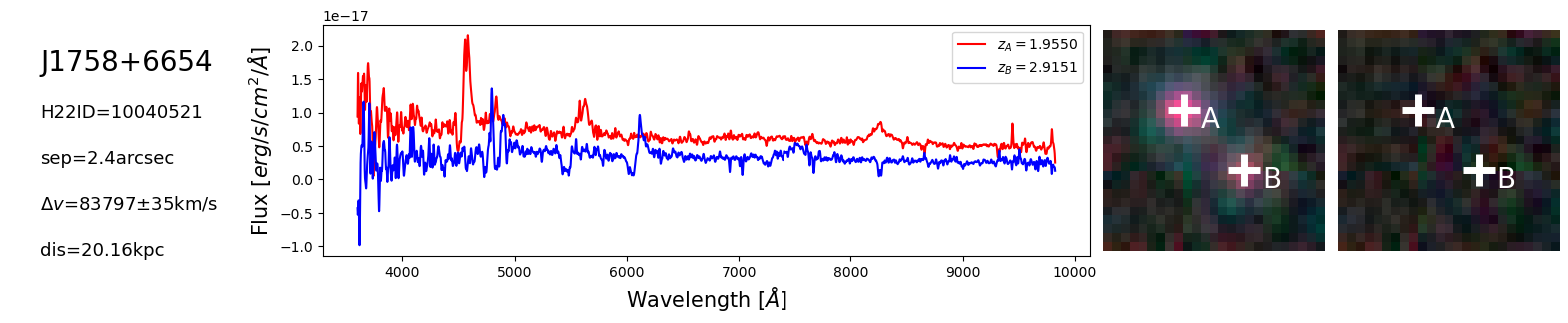}
    \caption{--Continued.}
\end{figure*}

\section{Summary and Discussion}\label{sec:diss}

In this study, we conducted follow-up spectroscopic observations of lensed-quasar candidates from the previously compiled H22 catalogue \citep{He2023}. These observations took place on October 15-16, 2023, at the Palomar Observatory in California, utilising the P200/DBSP instrument. Additionally, we cross-checked all H22 candidates with publicly available spectroscopic surveys, including DESI-EDR and SDSS. Through the combined efforts of \LIS, DBSP/P200, and existing spectroscopic surveys, we successfully confirmed two lensed quasars, six dual quasars, and eleven projected quasar pairs.

\begin{itemize}
    \item Strongly Lensed Quasars: J0746+1344 and J2121-0826 are two lensed quasars at $z_s$ = 3.105 and $z_s$ = 2.395. Assuming SIE mass model, the $\theta_E$ of the lensing system are 1$\arcsec$.208 and 0$\arcsec$.749, respectively. Both lenses were confirmed through DBSP spectroscopy.
    \item Dual Quasars: We identified six dual quasars, with two confirmed by DBSP spectra and the remainder from public datasets. The mean redshift of these pairs is 1.96, with J1929+6009 at $z$ = 3.28 being the highest redshift pair and J1643+3156 at $z$ = 0.59 the lowest. The projected physical separations range from 15.44 to 22.54 kpc.
    \item Projected Quasar Pairs: Eleven projected quasar pairs were confirmed, two through DBSP spectra and the rest from public datasets. The projected physical separations of these pairs, ranging from 10.96 to 39.07 kpc, make them suitable for studying the CGM of the lower redshift quasars\citep{Lau2018,Chen2023}.
    \item Potential Lensed Quasars: Based on DBSP spectra and \LIS, we identified three potential lensed quasars. These candidates require deeper imaging or resolved spectra for each image to conclusively determine their lensing nature.
\end{itemize}
One of the confirmed lensed quasars, J0746+1344, exhibits a strong flux anomaly. The lensing galaxy is located next to the brightest image (image B), which is unusual since the opposite configuration is typically observed. According to \LIS, over a two-year baseline, the magnitude of image B is more than one magnitude brighter than that of image A in the $grz$ bands. This is likely due to an ongoing microlensing effect.

We note that J0746+1344 lies within the WFST wide-file survey footprint \citep{Chen2023}, making it ideal candidate for time delay measurement, provided its time delay is less than 90 days. Both of the confirmed lensed quasars can be monitored by the Muztagh-Ata 1.9-meter Synergy Telescope \citep[MOST, see e.g., ][]{Zhu2023} in the future.
Projected quasar pairs serve as exceptional cosmic probes, playing a crucial role in studying the CGM of quasar hosts and their environments. Among our discoveries, J0422+0047 stands out as a particularly intriguing system. It offers the unique opportunity to simultaneously explore the CGM in both the outskirts of the quasar host and a coincidentally overlapped foreground galaxy. This rare configuration provides valuable insights into the diverse environments of the CGM, highlighting differences between AGNs and quiescent elliptical galaxies.
For the potential lensed quasars, deeper and higher resolution imaging can be anticipated from future missions such as Euclid \citep{Euclid-intro,Euclid2024}, Chinese Space Station Telescope \citep[CSST;][]{CSST-intro}, and Roman \citep{Roman-intro}. If confirmed, although their individual light curves cannot be resolved or separately recovered by WFST or MOST, time-delay measurements can still be conducted using blended light curves \citep{Shu2022}.

\begin{figure*}
    \centering
    \includegraphics[scale=0.5]{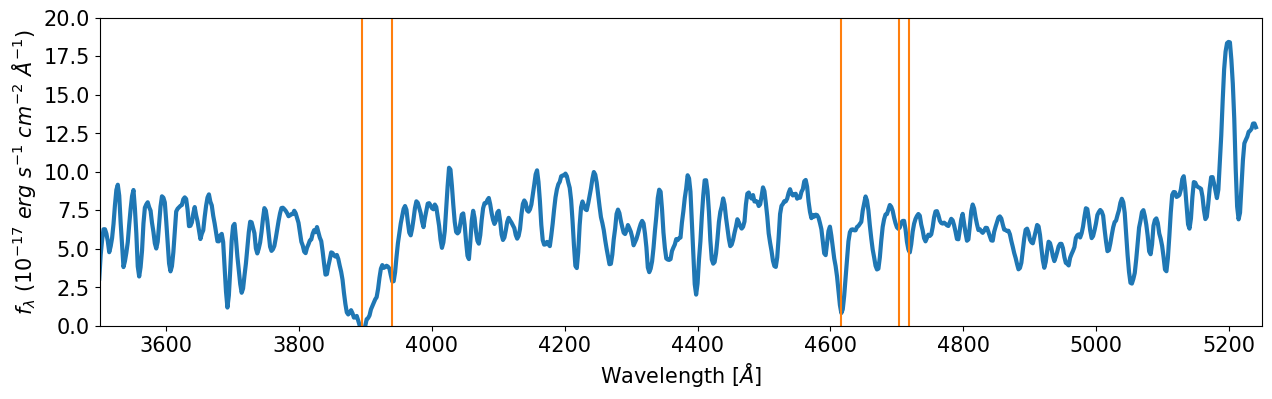}
    \caption{The blue side spectrum of J1929+6009B spans from 3500\AA to 5250\AA. The solid orange line represents the absorptions at \z{2.797}. From left to right, the features are $Ly_{\beta}$, $O_{VI}$, $Ly_{\alpha}$, and $N_{V}$ doublet.}
    \label{fig:J1929-abs}
\end{figure*}

In addition, we would like to emphasise an intriguing dual quasar discovered by P200, J1929+6009 at \z{3.28}. The redshift difference between the two components is less than 0.0001, and the projected separation is 19.28 kpc. In \cite{Hennawi2010}, 24 dual quasars were identified within a redshift range of 3 to 4, with a mean projected separation of 267 kpc. This highlights the uniqueness of the J1929+6009 system: its projected separation of 19.28 kpc at \z{3.28} is notably small.
Another interesting aspect is numerous absorptions span the range of 3500 \AA to 5200 \AA\ have been observed on the blue side of J1929+6009B, which are not visible in J1929+6009A, although they are only separated by \T{1.76} and at the same redshift. As illustrated in Figure \ref{fig:J1929-abs}, the most prominent feature appears to be a $Ly_{\beta}$ absorption at \z{2.797}. However, no counterpart can be found a such a redshift in the public available spectra data-sets.

\begin{acknowledgements}
This research uses data obtained through the Telescope Access Program (TAP), which has been funded by the TAP association, including Centre for Astronomical Mega-Science CAS(CAMS), XMU, PKU, THU, USTC, NJU, YNU, and SYSU. We thank the anonymous referee for very valuable and constructive comments that help us to improve the paper significantly. We thank Jiao Li, Chao Liu for insightful discussions. We thank \textbf{\textit{astropy}}, \textbf{\textit{pandas}}, \textbf{\textit{matplotlib}},  \textbf{\textit{astroscrappy}}, \textit{lenstronomy} for providing convenient and reliable python packages.\\

Z.H. acknowledges support from the China Postdoctoral Science Foundation under Grant Number GZC20232990 and the National Natural Science Foundation of China (Grant No. 12403104). R.L. acknowledges the support of the National Nature Science Foundation of China (No 12203050). Y.S. acknowledges the support from the National Science Foundation of China (12333001) and the China Manned Space Program through its Space Application System. G.L. acknowledges the support of the China Manned Spaced Project (CMS-CSST-2021-A12). N.L. acknowledges the support of the CAS Project for Young Scientists in Basic Research (No. YSBR-062), the science research grants from the China Manned Space Project (No. CMS-CSST-2021-A01), and the hospitality of the International Centre of Supernovae (ICESUN), Yunnan Key Laboratory at Yunnan Observatories, Chinese Academy of Sciences. D.D.S. acknowledges the support from the National Science Foundation of China (12303015) and the National Science Foundation of Jiangsu Province (BK20231106).\\

This project used data obtained with the Dark Energy Camera (DECam), which was constructed by the DES collaboration. Funding for the DES Projects has been provided by the U.S. Department of Energy, the U.S. National Science Foundation, the Ministry of Science and Education of Spain, the Science and Technology Facilities Council of the United Kingdom, the Higher Education Funding Council for England, the National Center for Supercomputing Applications at the University of Illinois at Urbana-Champaign, the Kavli Institute of Cosmological Physics at the University of Chicago, Center for Cosmology and Astro-Particle Physics at the Ohio State University, the Mitchell Institute for Fundamental Physics and Astronomy at Texas A\&M University, Financiadora de Estudos e Projetos, Fundacao Carlos Chagas Filho de Amparo, Financiadora de Estudos e Projetos, Fundacao Carlos Chagas Filho de Amparo a Pesquisa do Estado do Rio de Janeiro, Conselho Nacional de Desenvolvimento Cientifico e Tecnologico and the Ministerio da Ciencia, Tecnologia e Inovacao, the Deutsche Forschungsgemeinschaft and the Collaborating Institutions in the Dark Energy Survey. The Collaborating Institutions are Argonne National Laboratory, the University of California at Santa Cruz, the University of Cambridge, Centro de Investigaciones Energeticas, Medioambientales y Tecnologicas-Madrid, the University of Chicago, University College London, the DES-Brazil Consortium, the University of Edinburgh, the Eidgenossische Technische Hochschule (ETH) Zurich, Fermi National Accelerator Laboratory, the University of Illinois at Urbana-Champaign, the Institut de Ciencies de l’Espai (IEEC/CSIC), the Institut de Fisica d’Altes Energies, Lawrence Berkeley National Laboratory, the Ludwig Maximilians Universitat Munchen and the associated Excellence Cluster Universe, the University of Michigan, NSF’s NOIRLab, the University of Nottingham, the Ohio State University, the University of Pennsylvania, the University of Portsmouth, SLAC National Accelerator Laboratory, Stanford University, the University of Sussex, and Texas A\&M University.

\end{acknowledgements}

%
%

\bibliographystyle{aa}
\bibliography{cite-database}

\begin{thebibliography}{56}
\expandafter\ifx\csname natexlab\endcsname\relax\def\natexlab#1{#1}\fi

\bibitem[{{Anguita} {et~al.}(2008){Anguita}, {Schmidt}, {Turner}, {Wambsganss}, {Webster}, {Loomis}, {Long}, \& {McMillan}}]{Anguita2008}
{Anguita}, T., {Schmidt}, R.~W., {Turner}, E.~L., {et~al.} 2008, \aap, 480, 327

\bibitem[{{Astropy Collaboration} {et~al.}(2022){Astropy Collaboration}, {Price-Whelan}, {Lim}, {Earl}, {Starkman}, {Bradley}, {Shupe}, {Patil}, {Corrales}, {Brasseur}, {N{\"o}the}, {Donath}, {Tollerud}, {Morris}, {Ginsburg}, {Vaher}, {Weaver}, {Tocknell}, {Jamieson}, {van Kerkwijk}, {Robitaille}, {Merry}, {Bachetti}, {G{\"u}nther}, {Aldcroft}, {Alvarado-Montes}, {Archibald}, {B{\'o}di}, {Bapat}, {Barentsen}, {Baz{\'a}n}, {Biswas}, {Boquien}, {Burke}, {Cara}, {Cara}, {Conroy}, {Conseil}, {Craig}, {Cross}, {Cruz}, {D'Eugenio}, {Dencheva}, {Devillepoix}, {Dietrich}, {Eigenbrot}, {Erben}, {Ferreira}, {Foreman-Mackey}, {Fox}, {Freij}, {Garg}, {Geda}, {Glattly}, {Gondhalekar}, {Gordon}, {Grant}, {Greenfield}, {Groener}, {Guest}, {Gurovich}, {Handberg}, {Hart}, {Hatfield-Dodds}, {Homeier}, {Hosseinzadeh}, {Jenness}, {Jones}, {Joseph}, {Kalmbach}, {Karamehmetoglu}, {Ka{\l}uszy{\'n}ski}, {Kelley}, {Kern}, {Kerzendorf}, {Koch}, {Kulumani}, {Lee}, {Ly}, {Ma}, {MacBride}, {Maljaars}, {Muna}, {Murphy}, {Norman},
  {O'Steen}, {Oman}, {Pacifici}, {Pascual}, {Pascual-Granado}, {Patil}, {Perren}, {Pickering}, {Rastogi}, {Roulston}, {Ryan}, {Rykoff}, {Sabater}, {Sakurikar}, {Salgado}, {Sanghi}, {Saunders}, {Savchenko}, {Schwardt}, {Seifert-Eckert}, {Shih}, {Jain}, {Shukla}, {Sick}, {Simpson}, {Singanamalla}, {Singer}, {Singhal}, {Sinha}, {Sip{\H{o}}cz}, {Spitler}, {Stansby}, {Streicher}, {{\v{S}}umak}, {Swinbank}, {Taranu}, {Tewary}, {Tremblay}, {de Val-Borro}, {Van Kooten}, {Vasovi{\'c}}, {Verma}, {de Miranda Cardoso}, {Williams}, {Wilson}, {Winkel}, {Wood-Vasey}, {Xue}, {Yoachim}, {Zhang}, {Zonca}, \& {Astropy Project Contributors}}]{AstropyCollaboration2022}
{Astropy Collaboration}, {Price-Whelan}, A.~M., {Lim}, P.~L., {et~al.} 2022, \apj, 935, 167

\bibitem[{{Blanton} {et~al.}(2017){Blanton}, {Bershady}, {Abolfathi}, {Albareti}, {Allende Prieto}, {Almeida}, {Alonso-Garc{\'\i}a}, {Anders}, {Anderson}, {Andrews}, {Aquino-Ort{\'\i}z}, {Arag{\'o}n-Salamanca}, {Argudo-Fern{\'a}ndez}, {Armengaud}, {Aubourg}, {Avila-Reese}, {Badenes}, {Bailey}, {Barger}, {Barrera-Ballesteros}, {Bartosz}, {Bates}, {Baumgarten}, {Bautista}, {Beaton}, {Beers}, {Belfiore}, {Bender}, {Berlind}, {Bernardi}, {Beutler}, {Bird}, {Bizyaev}, {Blanc}, {Blomqvist}, {Bolton}, {Boquien}, {Borissova}, {van den Bosch}, {Bovy}, {Brandt}, {Brinkmann}, {Brownstein}, {Bundy}, {Burgasser}, {Burtin}, {Busca}, {Cappellari}, {Delgado Carigi}, {Carlberg}, {Carnero Rosell}, {Carrera}, {Chanover}, {Cherinka}, {Cheung}, {G{\'o}mez Maqueo Chew}, {Chiappini}, {Choi}, {Chojnowski}, {Chuang}, {Chung}, {Cirolini}, {Clerc}, {Cohen}, {Comparat}, {da Costa}, {Cousinou}, {Covey}, {Crane}, {Croft}, {Cruz-Gonzalez}, {Garrido Cuadra}, {Cunha}, {Damke}, {Darling}, {Davies}, {Dawson}, {de la Macorra}, {Dell'Agli}, {De
  Lee}, {Delubac}, {Di Mille}, {Diamond-Stanic}, {Cano-D{\'\i}az}, {Donor}, {Downes}, {Drory}, {du Mas des Bourboux}, {Duckworth}, {Dwelly}, {Dyer}, {Ebelke}, {Eigenbrot}, {Eisenstein}, {Emsellem}, {Eracleous}, {Escoffier}, {Evans}, {Fan}, {Fern{\'a}ndez-Alvar}, {Fernandez-Trincado}, {Feuillet}, {Finoguenov}, {Fleming}, {Font-Ribera}, {Fredrickson}, {Freischlad}, {Frinchaboy}, {Fuentes}, {Galbany}, {Garcia-Dias}, {Garc{\'\i}a-Hern{\'a}ndez}, {Gaulme}, {Geisler}, {Gelfand}, {Gil-Mar{\'\i}n}, {Gillespie}, {Goddard}, {Gonzalez-Perez}, {Grabowski}, {Green}, {Grier}, {Gunn}, {Guo}, {Guy}, {Hagen}, {Hahn}, {Hall}, {Harding}, {Hasselquist}, {Hawley}, {Hearty}, {Gonzalez Hern{\'a}ndez}, {Ho}, {Hogg}, {Holley-Bockelmann}, {Holtzman}, {Holzer}, {Huehnerhoff}, {Hutchinson}, {Hwang}, {Ibarra-Medel}, {da Silva Ilha}, {Ivans}, {Ivory}, {Jackson}, {Jensen}, {Johnson}, {Jones}, {J{\"o}nsson}, {Jullo}, {Kamble}, {Kinemuchi}, {Kirkby}, {Kitaura}, {Klaene}, {Knapp}, {Kneib}, {Kollmeier}, {Lacerna}, {Lane}, {Lang}, {Law},
  {Lazarz}, {Lee}, {Le Goff}, {Liang}, {Li}, {Li}, {Lian}, {Lima}, {Lin}, {Lin}, {Bertran de Lis}, {Liu}, {de Icaza Lizaola}, {Long}, {Lucatello}, {Lundgren}, {MacDonald}, {Deconto Machado}, {MacLeod}, {Mahadevan}, {Geimba Maia}, {Maiolino}, {Majewski}, {Malanushenko}, {Malanushenko}, {Manchado}, {Mao}, {Maraston}, {Marques-Chaves}, {Masseron}, {Masters}, {McBride}, {McDermid}, {McGrath}, {McGreer}, {Medina Pe{\~n}a}, {Melendez}, {Merloni}, {Merrifield}, {Meszaros}, {Meza}, {Minchev}, {Minniti}, {Miyaji}, {More}, {Mulchaey}, {M{\"u}ller-S{\'a}nchez}, {Muna}, {Munoz}, {Myers}, {Nair}, {Nandra}, {Correa do Nascimento}, {Negrete}, {Ness}, {Newman}, {Nichol}, {Nidever}, {Nitschelm}, {Ntelis}, {O'Connell}, {Oelkers}, {Oravetz}, {Oravetz}, {Pace}, {Padilla}, {Palanque-Delabrouille}, {Alonso Palicio}, {Pan}, {Parejko}, {Parikh}, {P{\^a}ris}, {Park}, {Patten}, {Peirani}, {Pellejero-Ibanez}, {Penny}, {Percival}, {Perez-Fournon}, {Petitjean}, {Pieri}, {Pinsonneault}, {Pisani}, {Poleski}, {Prada}, {Prakash}, {Queiroz},
  {Raddick}, {Raichoor}, {Barboza Rembold}, {Richstein}, {Riffel}, {Riffel}, {Rix}, {Robin}, {Rockosi}, {Rodr{\'\i}guez-Torres}, {Roman-Lopes}, {Rom{\'a}n-Z{\'u}{\~n}iga}, {Rosado}, {Ross}, {Rossi}, {Ruan}, {Ruggeri}, {Rykoff}, {Salazar-Albornoz}, {Salvato}, {S{\'a}nchez}, {Aguado}, {S{\'a}nchez-Gallego}, {Santana}, {Santiago}, {Sayres}, {Schiavon}, {da Silva Schimoia}, {Schlafly}, {Schlegel}, {Schneider}, {Schultheis}, {Schuster}, {Schwope}, {Seo}, {Shao}, {Shen}, {Shetrone}, {Shull}, {Simon}, {Skinner}, {Skrutskie}, {Slosar}, {Smith}, {Sobeck}, {Sobreira}, {Somers}, {Souto}, {Stark}, {Stassun}, {Stauffer}, {Steinmetz}, {Storchi-Bergmann}, {Streblyanska}, {Stringfellow}, {Su{\'a}rez}, {Sun}, {Suzuki}, {Szigeti}, {Taghizadeh-Popp}, {Tang}, {Tao}, {Tayar}, {Tembe}, {Teske}, {Thakar}, {Thomas}, {Thompson}, {Tinker}, {Tissera}, {Tojeiro}, {Hernandez Toledo}, {de la Torre}, {Tremonti}, {Troup}, {Valenzuela}, {Martinez Valpuesta}, {Vargas-Gonz{\'a}lez}, {Vargas-Maga{\~n}a}, {Vazquez}, {Villanova}, {Vivek}, {Vogt},
  {Wake}, {Walterbos}, {Wang}, {Weaver}, {Weijmans}, {Weinberg}, {Westfall}, {Whelan}, {Wild}, {Wilson}, {Wood-Vasey}, {Wylezalek}, {Xiao}, {Yan}, {Yang}, {Ybarra}, {Y{\`e}che}, {Zakamska}, {Zamora}, {Zarrouk}, {Zasowski}, {Zhang}, {Zhao}, {Zheng}, {Zheng}, {Zhou}, {Zhou}, {Zhu}, {Zoccali}, \& {Zou}}]{Blanton2017}
{Blanton}, M.~R., {Bershady}, M.~A., {Abolfathi}, B., {et~al.} 2017, \aj, 154, 28

\bibitem[{{Bolton} {et~al.}(2012){Bolton}, {Brownstein}, {Kochanek}, {Shu}, {Schlegel}, {Eisenstein}, {Wake}, {Connolly}, {Maraston}, {Arneson}, \& {Weaver}}]{Bolton12}
{Bolton}, A.~S., {Brownstein}, J.~R., {Kochanek}, C.~S., {et~al.} 2012, \apj, 757, 82

\bibitem[{{Boylan-Kolchin} {et~al.}(2008){Boylan-Kolchin}, {Ma}, \& {Quataert}}]{BoylanKolchin2008}
{Boylan-Kolchin}, M., {Ma}, C.-P., \& {Quataert}, E. 2008, \mnras, 383, 93

\bibitem[{{Braibant} {et~al.}(2014){Braibant}, {Hutsem{\'e}kers}, {Sluse}, {Anguita}, \& {Garc{\'\i}a-Vergara}}]{Braibant2014}
{Braibant}, L., {Hutsem{\'e}kers}, D., {Sluse}, D., {Anguita}, T., \& {Garc{\'\i}a-Vergara}, C.~J. 2014, \aap, 565, L11

\bibitem[{{Cai} {et~al.}(2019){Cai}, {Cantalupo}, {Prochaska}, {Arrigoni Battaia}, {Burchett}, {Li}, {Chisholm}, {Bundy}, \& {Hennawi}}]{Cai2019}
{Cai}, Z., {Cantalupo}, S., {Prochaska}, J.~X., {et~al.} 2019, \apjs, 245, 23

\bibitem[{{Cao} {et~al.}(2023){Cao}, {Li}, {Li}, {Li}, {Chen}, {Ding}, {Shan}, {Hu}, {Zhan}, {Xing}, {Zhan}, {Wei}, {Du}, \& {Cao}}]{Cao2023}
{Cao}, X., {Li}, R., {Li}, N., {et~al.} 2023, arXiv e-prints, arXiv:2312.06239

\bibitem[{{Cao} {et~al.}(2018){Cao}, {Gong}, {Meng}, {Xu}, {Chen}, {Guo}, {Li}, {Liu}, {Xue}, {Cao}, {Fu}, {Zhang}, {Wang}, \& {Zhan}}]{CSST-intro}
{Cao}, Y., {Gong}, Y., {Meng}, X.-M., {et~al.} 2018, \mnras, 480, 2178

\bibitem[{{Chambers} {et~al.}(2016){Chambers}, {Magnier}, {Metcalfe}, {Flewelling}, {Huber}, {Waters}, {Denneau}, {Draper}, {Farrow}, {Finkbeiner}, {Holmberg}, {Koppenhoefer}, {Price}, {Rest}, {Saglia}, {Schlafly}, {Smartt}, {Sweeney}, {Wainscoat}, {Burgett}, {Chastel}, {Grav}, {Heasley}, {Hodapp}, {Jedicke}, {Kaiser}, {Kudritzki}, {Luppino}, {Lupton}, {Monet}, {Morgan}, {Onaka}, {Shiao}, {Stubbs}, {Tonry}, {White}, {Ba{\~n}ados}, {Bell}, {Bender}, {Bernard}, {Boegner}, {Boffi}, {Botticella}, {Calamida}, {Casertano}, {Chen}, {Chen}, {Cole}, {Deacon}, {Frenk}, {Fitzsimmons}, {Gezari}, {Gibbs}, {Goessl}, {Goggia}, {Gourgue}, {Goldman}, {Grant}, {Grebel}, {Hambly}, {Hasinger}, {Heavens}, {Heckman}, {Henderson}, {Henning}, {Holman}, {Hopp}, {Ip}, {Isani}, {Jackson}, {Keyes}, {Koekemoer}, {Kotak}, {Le}, {Liska}, {Long}, {Lucey}, {Liu}, {Martin}, {Masci}, {McLean}, {Mindel}, {Misra}, {Morganson}, {Murphy}, {Obaika}, {Narayan}, {Nieto-Santisteban}, {Norberg}, {Peacock}, {Pier}, {Postman}, {Primak}, {Rae}, {Rai},
  {Riess}, {Riffeser}, {Rix}, {R{\"o}ser}, {Russel}, {Rutz}, {Schilbach}, {Schultz}, {Scolnic}, {Strolger}, {Szalay}, {Seitz}, {Small}, {Smith}, {Soderblom}, {Taylor}, {Thomson}, {Taylor}, {Thakar}, {Thiel}, {Thilker}, {Unger}, {Urata}, {Valenti}, {Wagner}, {Walder}, {Walter}, {Watters}, {Werner}, {Wood-Vasey}, \& {Wyse}}]{Panstarss2016}
{Chambers}, K.~C., {Magnier}, E.~A., {Metcalfe}, N., {et~al.} 2016, arXiv e-prints, arXiv:1612.05560

\bibitem[{{Chen} {et~al.}(2023){Chen}, {Qin}, {Cai}, {Zhou}, {Chen}, {Pang}, {Wang}, \& {Cheng}}]{Chen2023}
{Chen}, Z.-F., {Qin}, H.-C., {Cai}, J.-T., {et~al.} 2023, \apjs, 265, 46

\bibitem[{{Dawes} {et~al.}(2023){Dawes}, {Storfer}, {Huang}, {Aldering}, {Cikota}, {Dey}, \& {Schlegel}}]{Dawes2023}
{Dawes}, C., {Storfer}, C., {Huang}, X., {et~al.} 2023, \apjs, 269, 61

\bibitem[{{de Jong} {et~al.}(2019){de Jong}, {Agertz}, {Berbel}, {Aird}, {Alexander}, {Amarsi}, {Anders}, {Andrae}, {Ansarinejad}, {Ansorge}, {Antilogus}, {Anwand-Heerwart}, {Arentsen}, {Arnadottir}, {Asplund}, {Auger}, {Azais}, {Baade}, {Baker}, {Baker}, {Balbinot}, {Baldry}, {Banerji}, {Barden}, {Barklem}, {Barth{\'e}l{\'e}my-Mazot}, {Battistini}, {Bauer}, {Bell}, {Bellido-Tirado}, {Bellstedt}, {Belokurov}, {Bensby}, {Bergemann}, {Bestenlehner}, {Bielby}, {Bilicki}, {Blake}, {Bland-Hawthorn}, {Boeche}, {Boland}, {Boller}, {Bongard}, {Bongiorno}, {Bonifacio}, {Boudon}, {Brooks}, {Brown}, {Brown}, {Br{\"u}ggen}, {Brynnel}, {Brzeski}, {Buchert}, {Buschkamp}, {Caffau}, {Caillier}, {Carrick}, {Casagrande}, {Case}, {Casey}, {Cesarini}, {Cescutti}, {Chapuis}, {Chiappini}, {Childress}, {Christlieb}, {Church}, {Cioni}, {Cluver}, {Colless}, {Collett}, {Comparat}, {Cooper}, {Couch}, {Courbin}, {Croom}, {Croton}, {Daguis{\'e}}, {Dalton}, {Davies}, {Davis}, {de Laverny}, {Deason}, {Dionies}, {Disseau}, {Doel},
  {D{\"o}scher}, {Driver}, {Dwelly}, {Eckert}, {Edge}, {Edvardsson}, {Youssoufi}, {Elhaddad}, {Enke}, {Erfanianfar}, {Farrell}, {Fechner}, {Feiz}, {Feltzing}, {Ferreras}, {Feuerstein}, {Feuillet}, {Finoguenov}, {Ford}, {Fotopoulou}, {Fouesneau}, {Frenk}, {Frey}, {Gaessler}, {Geier}, {Gentile Fusillo}, {Gerhard}, {Giannantonio}, {Giannone}, {Gibson}, {Gillingham}, {Gonz{\'a}lez-Fern{\'a}ndez}, {Gonzalez-Solares}, {Gottloeber}, {Gould}, {Grebel}, {Gueguen}, {Guiglion}, {Haehnelt}, {Hahn}, {Hansen}, {Hartman}, {Hauptner}, {Hawkins}, {Haynes}, {Haynes}, {Heiter}, {Helmi}, {Aguayo}, {Hewett}, {Hinton}, {Hobbs}, {Hoenig}, {Hofman}, {Hook}, {Hopgood}, {Hopkins}, {Hourihane}, {Howes}, {Howlett}, {Huet}, {Irwin}, {Iwert}, {Jablonka}, {Jahn}, {Jahnke}, {Jarno}, {Jin}, {Jofre}, {Johl}, {Jones}, {J{\"o}nsson}, {Jordan}, {Karovicova}, {Khalatyan}, {Kelz}, {Kennicutt}, {King}, {Kitaura}, {Klar}, {Klauser}, {Kneib}, {Koch}, {Koposov}, {Kordopatis}, {Korn}, {Kosmalski}, {Kotak}, {Kovalev}, {Kreckel}, {Kripak}, {Krumpe},
  {Kuijken}, {Kunder}, {Kushniruk}, {Lam}, {Lamer}, {Laurent}, {Lawrence}, {Lehmitz}, {Lemasle}, {Lewis}, {Li}, {Lidman}, {Lind}, {Liske}, {Lizon}, {Loveday}, {Ludwig}, {McDermid}, {Maguire}, {Mainieri}, {Mali}, {Mandel}, {Mandel}, {Mannering}, {Martell}, {Martinez Delgado}, {Matijevic}, {McGregor}, {McMahon}, {McMillan}, {Mena}, {Merloni}, {Meyer}, {Michel}, {Micheva}, {Migniau}, {Minchev}, {Monari}, {Muller}, {Murphy}, {Muthukrishna}, {Nandra}, {Navarro}, {Ness}, {Nichani}, {Nichol}, {Nicklas}, {Niederhofer}, {Norberg}, {Obreschkow}, {Oliver}, {Owers}, {Pai}, {Pankratow}, {Parkinson}, {Paschke}, {Paterson}, {Pecontal}, {Parry}, {Phillips}, {Pillepich}, {Pinard}, {Pirard}, {Piskunov}, {Plank}, {Pl{\"u}schke}, {Pons}, {Popesso}, {Power}, {Pragt}, {Pramskiy}, {Pryer}, {Quattri}, {Queiroz}, {Quirrenbach}, {Rahurkar}, {Raichoor}, {Ramstedt}, {Rau}, {Recio-Blanco}, {Reiss}, {Renaud}, {Revaz}, {Rhode}, {Richard}, {Richter}, {Rix}, {Robotham}, {Roelfsema}, {Romaniello}, {Rosario}, {Rothmaier}, {Roukema}, {Ruchti},
  {Rupprecht}, {Rybizki}, {Ryde}, {Saar}, {Sadler}, {Sahl{\'e}n}, {Salvato}, {Sassolas}, {Saunders}, {Saviauk}, {Sbordone}, {Schmidt}, {Schnurr}, {Scholz}, {Schwope}, {Seifert}, {Shanks}, {Sheinis}, {Sivov}, {Sk{\'u}lad{\'o}ttir}, {Smartt}, {Smedley}, {Smith}, {Smith}, {Sorce}, {Spitler}, {Starkenburg}, {Steinmetz}, {Stilz}, {Storm}, {Sullivan}, {Sutherland}, {Swann}, {Tamone}, {Taylor}, {Teillon}, {Tempel}, {ter Horst}, {Thi}, {Tolstoy}, {Trager}, {Traven}, {Tremblay}, {Tresse}, {Valentini}, {van de Weygaert}, {van den Ancker}, {Veljanoski}, {Venkatesan}, {Wagner}, {Wagner}, {Walcher}, {Waller}, {Walton}, {Wang}, {Winkler}, {Wisotzki}, {Worley}, {Worseck}, {Xiang}, {Xu}, {Yong}, {Zhao}, {Zheng}, {Zscheyge}, \& {Zucker}}]{deJong2019}
{de Jong}, R.~S., {Agertz}, O., {Berbel}, A.~A., {et~al.} 2019, The Messenger, 175, 3

\bibitem[{{De Rosa} {et~al.}(2019){De Rosa}, {Vignali}, {Bogdanovi{\'c}}, {Capelo}, {Charisi}, {Dotti}, {Husemann}, {Lusso}, {Mayer}, {Paragi}, {Runnoe}, {Sesana}, {Steinborn}, {Bianchi}, {Colpi}, {del Valle}, {Frey}, {Gab{\'a}nyi}, {Giustini}, {Guainazzi}, {Haiman}, {Herrera Ruiz}, {Herrero-Illana}, {Iwasawa}, {Komossa}, {Lena}, {Loiseau}, {Perez-Torres}, {Piconcelli}, \& {Volonteri}}]{DeRosa2019}
{De Rosa}, A., {Vignali}, C., {Bogdanovi{\'c}}, T., {et~al.} 2019, \nar, 86, 101525

\bibitem[{{DESI Collaboration} {et~al.}(2016){DESI Collaboration}, {Aghamousa}, {Aguilar}, {Ahlen}, {Alam}, {Allen}, {Allende Prieto}, {Annis}, {Bailey}, {Balland}, {Ballester}, {Baltay}, {Beaufore}, {Bebek}, {Beers}, {Bell}, {Bernal}, {Besuner}, {Beutler}, {Blake}, {Bleuler}, {Blomqvist}, {Blum}, {Bolton}, {Briceno}, {Brooks}, {Brownstein}, {Buckley-Geer}, {Burden}, {Burtin}, {Busca}, {Cahn}, {Cai}, {Cardiel-Sas}, {Carlberg}, {Carton}, {Casas}, {Castander}, {Cervantes-Cota}, {Claybaugh}, {Close}, {Coker}, {Cole}, {Comparat}, {Cooper}, {Cousinou}, {Crocce}, {Cuby}, {Cunningham}, {Davis}, {Dawson}, {de la Macorra}, {De Vicente}, {Delubac}, {Derwent}, {Dey}, {Dhungana}, {Ding}, {Doel}, {Duan}, {Ealet}, {Edelstein}, {Eftekharzadeh}, {Eisenstein}, {Elliott}, {Escoffier}, {Evatt}, {Fagrelius}, {Fan}, {Fanning}, {Farahi}, {Farihi}, {Favole}, {Feng}, {Fernandez}, {Findlay}, {Finkbeiner}, {Fitzpatrick}, {Flaugher}, {Flender}, {Font-Ribera}, {Forero-Romero}, {Fosalba}, {Frenk}, {Fumagalli}, {Gaensicke}, {Gallo},
  {Garcia-Bellido}, {Gaztanaga}, {Pietro Gentile Fusillo}, {Gerard}, {Gershkovich}, {Giannantonio}, {Gillet}, {Gonzalez-de-Rivera}, {Gonzalez-Perez}, {Gott}, {Graur}, {Gutierrez}, {Guy}, {Habib}, {Heetderks}, {Heetderks}, {Heitmann}, {Hellwing}, {Herrera}, {Ho}, {Holland}, {Honscheid}, {Huff}, {Hutchinson}, {Huterer}, {Hwang}, {Illa Laguna}, {Ishikawa}, {Jacobs}, {Jeffrey}, {Jelinsky}, {Jennings}, {Jiang}, {Jimenez}, {Johnson}, {Joyce}, {Jullo}, {Juneau}, {Kama}, {Karcher}, {Karkar}, {Kehoe}, {Kennamer}, {Kent}, {Kilbinger}, {Kim}, {Kirkby}, {Kisner}, {Kitanidis}, {Kneib}, {Koposov}, {Kovacs}, {Koyama}, {Kremin}, {Kron}, {Kronig}, {Kueter-Young}, {Lacey}, {Lafever}, {Lahav}, {Lambert}, {Lampton}, {Landriau}, {Lang}, {Lauer}, {Le Goff}, {Le Guillou}, {Le Van Suu}, {Lee}, {Lee}, {Leitner}, {Lesser}, {Levi}, {L'Huillier}, {Li}, {Liang}, {Lin}, {Linder}, {Loebman}, {Luki{\'c}}, {Ma}, {MacCrann}, {Magneville}, {Makarem}, {Manera}, {Manser}, {Marshall}, {Martini}, {Massey}, {Matheson}, {McCauley}, {McDonald},
  {McGreer}, {Meisner}, {Metcalfe}, {Miller}, {Miquel}, {Moustakas}, {Myers}, {Naik}, {Newman}, {Nichol}, {Nicola}, {Nicolati da Costa}, {Nie}, {Niz}, {Norberg}, {Nord}, {Norman}, {Nugent}, {O'Brien}, {Oh}, {Olsen}, {Padilla}, {Padmanabhan}, {Padmanabhan}, {Palanque-Delabrouille}, {Palmese}, {Pappalardo}, {P{\^a}ris}, {Park}, {Patej}, {Peacock}, {Peiris}, {Peng}, {Percival}, {Perruchot}, {Pieri}, {Pogge}, {Pollack}, {Poppett}, {Prada}, {Prakash}, {Probst}, {Rabinowitz}, {Raichoor}, {Ree}, {Refregier}, {Regal}, {Reid}, {Reil}, {Rezaie}, {Rockosi}, {Roe}, {Ronayette}, {Roodman}, {Ross}, {Ross}, {Rossi}, {Rozo}, {Ruhlmann-Kleider}, {Rykoff}, {Sabiu}, {Samushia}, {Sanchez}, {Sanchez}, {Schlegel}, {Schneider}, {Schubnell}, {Secroun}, {Seljak}, {Seo}, {Serrano}, {Shafieloo}, {Shan}, {Sharples}, {Sholl}, {Shourt}, {Silber}, {Silva}, {Sirk}, {Slosar}, {Smith}, {Smoot}, {Som}, {Song}, {Sprayberry}, {Staten}, {Stefanik}, {Tarle}, {Sien Tie}, {Tinker}, {Tojeiro}, {Valdes}, {Valenzuela}, {Valluri}, {Vargas-Magana},
  {Verde}, {Walker}, {Wang}, {Wang}, {Weaver}, {Weaverdyck}, {Wechsler}, {Weinberg}, {White}, {Yang}, {Yeche}, {Zhang}, {Zhao}, {Zheng}, {Zhou}, {Zhou}, {Zhu}, {Zou}, \& {Zu}}]{DESICollaboration2016}
{DESI Collaboration}, {Aghamousa}, A., {Aguilar}, J., {et~al.} 2016, arXiv e-prints, arXiv:1611.00036

\bibitem[{{Dey} {et~al.}(2019){Dey}, {Schlegel}, {Lang}, {Blum}, {Burleigh}, {Fan}, {Findlay}, {Finkbeiner}, {Herrera}, {Juneau}, {Landriau}, {Levi}, {McGreer}, {Meisner}, {Myers}, {Moustakas}, {Nugent}, {Patej}, {Schlafly}, {Walker}, {Valdes}, {Weaver}, {Y{\`e}che}, {Zou}, {Zhou}, {Abareshi}, {Abbott}, {Abolfathi}, {Aguilera}, {Alam}, {Allen}, {Alvarez}, {Annis}, {Ansarinejad}, {Aubert}, {Beechert}, {Bell}, {BenZvi}, {Beutler}, {Bielby}, {Bolton}, {Brice{\~n}o}, {Buckley-Geer}, {Butler}, {Calamida}, {Carlberg}, {Carter}, {Casas}, {Castander}, {Choi}, {Comparat}, {Cukanovaite}, {Delubac}, {DeVries}, {Dey}, {Dhungana}, {Dickinson}, {Ding}, {Donaldson}, {Duan}, {Duckworth}, {Eftekharzadeh}, {Eisenstein}, {Etourneau}, {Fagrelius}, {Farihi}, {Fitzpatrick}, {Font-Ribera}, {Fulmer}, {G{\"a}nsicke}, {Gaztanaga}, {George}, {Gerdes}, {Gontcho}, {Gorgoni}, {Green}, {Guy}, {Harmer}, {Hernandez}, {Honscheid}, {Huang}, {James}, {Jannuzi}, {Jiang}, {Joyce}, {Karcher}, {Karkar}, {Kehoe}, {Kneib}, {Kueter-Young}, {Lan},
  {Lauer}, {Le Guillou}, {Le Van Suu}, {Lee}, {Lesser}, {Perreault Levasseur}, {Li}, {Mann}, {Marshall}, {Mart{\'\i}nez-V{\'a}zquez}, {Martini}, {du Mas des Bourboux}, {McManus}, {Meier}, {M{\'e}nard}, {Metcalfe}, {Mu{\~n}oz-Guti{\'e}rrez}, {Najita}, {Napier}, {Narayan}, {Newman}, {Nie}, {Nord}, {Norman}, {Olsen}, {Paat}, {Palanque-Delabrouille}, {Peng}, {Poppett}, {Poremba}, {Prakash}, {Rabinowitz}, {Raichoor}, {Rezaie}, {Robertson}, {Roe}, {Ross}, {Ross}, {Rudnick}, {Safonova}, {Saha}, {S{\'a}nchez}, {Savary}, {Schweiker}, {Scott}, {Seo}, {Shan}, {Silva}, {Slepian}, {Soto}, {Sprayberry}, {Staten}, {Stillman}, {Stupak}, {Summers}, {Sien Tie}, {Tirado}, {Vargas-Maga{\~n}a}, {Vivas}, {Wechsler}, {Williams}, {Yang}, {Yang}, {Yapici}, {Zaritsky}, {Zenteno}, {Zhang}, {Zhang}, {Zhou}, \& {Zhou}}]{Dey2019}
{Dey}, A., {Schlegel}, D.~J., {Lang}, D., {et~al.} 2019, \aj, 157, 168

\bibitem[{{Dux} {et~al.}(2024){Dux}, {Lemon}, {Courbin}, {Neira}, {Anguita}, {Galan}, {Kim}, {Hempel}, {Hempel}, \& {Lachaume}}]{Dux2024}
{Dux}, F., {Lemon}, C., {Courbin}, F., {et~al.} 2024, \aap, 682, A47

\bibitem[{{Dux} {et~al.}(2023){Dux}, {Lemon}, {Courbin}, {Sluse}, {Smette}, {Anguita}, \& {Neira}}]{Dux2023}
{Dux}, F., {Lemon}, C., {Courbin}, F., {et~al.} 2023, \aap, 679, L4

\bibitem[{{Eifler} {et~al.}(2021){Eifler}, {Miyatake}, {Krause}, {Heinrich}, {Miranda}, {Hirata}, {Xu}, {Hemmati}, {Simet}, {Capak}, {Choi}, {Dor{\'e}}, {Doux}, {Fang}, {Hounsell}, {Huff}, {Huang}, {Jarvis}, {Kruk}, {Masters}, {Rozo}, {Scolnic}, {Spergel}, {Troxel}, {von der Linden}, {Wang}, {Weinberg}, {Wenzl}, \& {Wu}}]{Roman-intro}
{Eifler}, T., {Miyatake}, H., {Krause}, E., {et~al.} 2021, \mnras, 507, 1746

\bibitem[{{Euclid Collaboration} {et~al.}(2024){Euclid Collaboration}, {Mellier}, {Abdurro'uf}, {Acevedo Barroso}, {Ach{\'u}carro}, {Adamek}, {Adam}, {Addison}, {Aghanim}, {Aguena}, {Ajani}, {Akrami}, {Al-Bahlawan}, {Alavi}, {Albuquerque}, {Alestas}, {Alguero}, {Allaoui}, {Allen}, {Allevato}, {Alonso-Tetilla}, {Altieri}, {Alvarez-Candal}, {Amara}, {Amendola}, {Amiaux}, {Andika}, {Andreon}, {Andrews}, {Angora}, {Angulo}, {Annibali}, {Anselmi}, {Anselmi}, {Arcari}, {Archidiacono}, {Aric{\`o}}, {Arnaud}, {Arnouts}, {Asgari}, {Asorey}, {Atayde}, {Atek}, {Atrio-Barandela}, {Aubert}, {Aubourg}, {Auphan}, {Auricchio}, {Aussel}, {Aussel}, {Avelino}, {Avgoustidis}, {Avila}, {Awan}, {Azzollini}, {Baccigalupi}, {Bachelet}, {Bacon}, {Baes}, {Bagley}, {Bahr-Kalus}, {Balaguera-Antolinez}, {Balbinot}, {Balcells}, {Baldi}, {Baldry}, {Balestra}, {Ballardini}, {Ballester}, {Balogh}, {Ba{\~n}ados}, {Barbier}, {Bardelli}, {Barreiro}, {Barriere}, {Barros}, {Barthelemy}, {Bartolo}, {Basset}, {Battaglia}, {Battisti}, {Baugh},
  {Baumont}, {Bazzanini}, {Beaulieu}, {Beckmann}, {Belikov}, {Bel}, {Bellagamba}, {Bella}, {Bellini}, {Benabed}, {Bender}, {Benevento}, {Bennett}, {Benson}, {Bergamini}, {Bermejo-Climent}, {Bernardeau}, {Bertacca}, {Berthe}, {Berthier}, {Bethermin}, {Beutler}, {Bevillon}, {Bhargava}, {Bhatawdekar}, {Bisigello}, {Biviano}, {Blake}, {Blanchard}, {Blazek}, {Blot}, {Bosco}, {Bodendorf}, {Boenke}, {B{\"o}hringer}, {Bolzonella}, {Bonchi}, {Bonici}, {Bonino}, {Bonino}, {Bonvin}, {Bon}, {Booth}, {Borgani}, {Borlaff}, {Borsato}, {Bosco}, {Bose}, {Botticella}, {Boucaud}, {Bouche}, {Boucher}, {Boutigny}, {Bouvard}, {Bouy}, {Bowler}, {Bozza}, {Bozzo}, {Branchini}, {Brau-Nogue}, {Brekke}, {Bremer}, {Brescia}, {Breton}, {Brinchmann}, {Brinckmann}, {Brockley-Blatt}, {Brodwin}, {Brouard}, {Brown}, {Bruton}, {Bucko}, {Buddelmeijer}, {Buenadicha}, {Buitrago}, {Burger}, {Burigana}, {Busillo}, {Busonero}, {Cabanac}, {Cabayol-Garcia}, {Cagliari}, {Caillat}, {Caillat}, {Calabrese}, {Calabro}, {Calderone}, {Calura}, {Camacho
  Quevedo}, {Camera}, {Campos}, {Canas-Herrera}, {Candini}, {Cantiello}, {Capobianco}, {Cappellaro}, {Cappelluti}, {Cappi}, {Caputi}, {Cara}, {Carbone}, {Cardone}, {Carella}, {Carlberg}, {Carle}, {Carminati}, {Caro}, {Carrasco}, {Carretero}, {Carrilho}, {Carron Duque}, {Carry}, {Carvalho}, {Carvalho}, {Casas}, {Casas}, {Casenove}, {Casey}, {Cassata}, {Castander}, {Castelao}, {Castellano}, {Castiblanco}, {Castignani}, {Castro}, {Cavet}, {Cavuoti}, {Chabaud}, {Chambers}, {Charles}, {Charlot}, {Chartab}, {Chary}, {Chaumeil}, {Cho}, {Chon}, {Ciancetta}, {Ciliegi}, {Cimatti}, {Cimino}, {Cioni}, {Claydon}, {Cleland}, {Cl{\'e}ment}, {Clements}, {Clerc}, {Clesse}, {Codis}, {Cogato}, {Colbert}, {Cole}, {Coles}, {Collett}, {Collins}, {Colodro-Conde}, {Colombo}, {Combes}, {Conforti}, {Congedo}, {Conseil}, {Conselice}, {Contarini}, {Contini}, {Conversi}, {Cooray}, {Copin}, {Corasaniti}, {Corcho-Caballero}, {Corcione}, {Cordes}, {Corpace}, {Correnti}, {Costanzi}, {Costille}, {Courbin}, {Courcoult Mifsud}, {Courtois},
  {Cousinou}, {Covone}, {Cowell}, {Cragg}, {Cresci}, {Cristiani}, {Crocce}, {Cropper}, {E Crouzet}, {Csizi}, {Cuby}, {Cucchetti}, {Cucciati}, {Cuillandre}, {Cunha}, {Cuozzo}, {Daddi}, {D'Addona}, {Dafonte}, {Dagoneau}, {Dalessandro}, {Dalton}, {D'Amico}, {Dannerbauer}, {Danto}, {Das}, {Da Silva}, {da Silva}, {Daste}, {Davies}, {Davini}, {de Boer}, {Decarli}, {De Caro}, {Degaudenzi}, {Degni}, {de Jong}, {de la Bella}, {de la Torre}, {Delhaise}, {Delley}, {Delucchi}, {De Lucia}, {Denniston}, {De Paolis}, {De Petris}, {Derosa}, {Desai}, {Desjacques}, {Despali}, {Desprez}, {De Vicente-Albendea}, {Deville}, {Dias}, {D{\'\i}az-S{\'a}nchez}, {Diaz}, {Di Domizio}, {Diego}, {Di Ferdinando}, {Di Giorgio}, {Dimauro}, {Dinis}, {Dolag}, {Dolding}, {Dole}, {Dom{\'\i}nguez S{\'a}nchez}, {Dor{\'e}}, {Dournac}, {Douspis}, {Dreihahn}, {Droge}, {Dryer}, {Dubath}, {Duc}, {Ducret}, {Duffy}, {Dufresne}, {Duncan}, {Dupac}, {Duret}, {Durrer}, {Durret}, {Dusini}, {Ealet}, {Eggemeier}, {Eisenhardt}, {Elbaz}, {Elkhashab}, {Ellien},
  {Endicott}, {Enia}, {Erben}, {Escartin Vigo}, {Escoffier}, {Escudero Sanz}, {Essert}, {Ettori}, {Ezziati}, {Fabbian}, {Fabricius}, {Fang}, {Farina}, {Farina}, {Farinelli}, {Farrens}, {Faustini}, {Feltre}, {Ferguson}, {Ferrando}, {Ferrari}, {Ferr{\'e}-Mateu}, {Ferreira}, {Ferreras}, {Ferrero}, {Ferriol}, {Ferruit}, {Filleul}, {Finelli}, {Finkelstein}, {Finoguenov}, {Fiorini}, {Flentge}, {Focardi}, {Fonseca}, {Fontana}, {Fontanot}, {Fornari}, {Fosalba}, {Fossati}, {Fotopoulou}, {Fouchez}, {Fourmanoit}, {Frailis}, {Fraix-Burnet}, {Franceschi}, {Franco}, {Franzetti}, {Freihoefer}, {Frittoli}, {Frugier}, {Frusciante}, {Fumagalli}, {Fumagalli}, {Fumana}, {Fu}, {Gabarra}, {Galeotta}, {Galluccio}, {Ganga}, {Gao}, {Garc{\'\i}a-Bellido}, {Garcia}, {Gardner}, {Garilli}, {Gaspar-Venancio}, {Gasparetto}, {Gautard}, {Gavazzi}, {Gaztanaga}, {Genolet}, {Genova Santos}, {Gentile}, {George}, {Ghaffari}, {Giacomini}, {Gianotti}, {Gibb}, {Gillard}, {Gillis}, {Ginolfi}, {Giocoli}, {Girardi}, {Giri}, {Goh}, {G{\'o}mez-Alvarez},
  {Gonzalez}, {Gonzalez}, {Gonzalez}, {Gouyou Beauchamps}, {Gozaliasl}, {Gracia-Carpio}, {Grandis}, {Granett}, {Granvik}, {Grazian}, {Gregorio}, {Grenet}, {Grillo}, {Grupp}, {Gruppioni}, {Gruppuso}, {Guerbuez}, {Guerrini}, {Guidi}, {Guillard}, {Gutierrez}, {Guttridge}, {Guzzo}, {Gwyn}, {Haapala}, {Haase}, {Haddow}, {Hailey}, {Hall}, {Hall}, {Hamaus}, {Haridasu}, {Harnois-D{\'e}raps}, {Harper}, {Hartley}, {Hasinger}, {Hassani}, {Hatch}, {Haugan}, {H{\"a}u{\ss}ler}, {Heavens}, {Heisenberg}, {Helmi}, {Helou}, {Hemmati}, {Henares}, {Herent}, {Hern{\'a}ndez-Monteagudo}, {Heuberger}, {Hewett}, {Heydenreich}, {Hildebrandt}, {Hirschmann}, {Hjorth}, {Hoar}, {Hoekstra}, {Holland}, {Holliman}, {Holmes}, {Hook}, {Horeau}, {Hormuth}, {Hornstrup}, {Hosseini}, {Hu}, {Hudelot}, {Hudson}, {Huertas-Company}, {Huff}, {Hughes}, {Humphrey}, {Hunt}, {Huynh}, {Ibata}, {Ichikawa}, {Iglesias-Groth}, {Ilbert}, {Ili{\'c}}, {Ingoglia}, {Iodice}, {Israel}, {Israelsson}, {Izzo}, {Jablonka}, {Jackson}, {Jacobson}, {Jafariyazani}, {Jahnke},
  {Jansen}, {Jarvis}, {Jasche}, {Jauzac}, {Jeffrey}, {Jhabvala}, {Jimenez-Teja}, {Jimenez Mu{\~n}oz}, {Joachimi}, {Johansson}, {Joudaki}, {Jullo}, {Kajava}, {Kang}, {Kannawadi}, {Kansal}, {Karagiannis}, {K{\"a}rcher}, {Kashlinsky}, {Kazandjian}, {Keck}, {Keih{\"a}nen}, {Kerins}, {Kermiche}, {Khalil}, {Kiessling}, {Kiiveri}, {Kilbinger}, {Kim}, {King}, {Kirkpatrick}, {Kitching}, {Kluge}, {Knabenhans}, {Knapen}, {Knebe}, {Kneib}, {Kohley}, {Koopmans}, {Koskinen}, {Koulouridis}, {Kou}, {Kov{\'a}cs}, {Kova\{{\v{c}}\}i{\'c}}, {Kowalczyk}, {Koyama}, {Kraljic}, {Krause}, {Kruk}, {Kubik}, {Kuchner}, {Kuijken}, {K{\"u}mmel}, {Kunz}, {Kurki-Suonio}, {Lacasa}, {Lacey}, {La Franca}, {Lagarde}, {Lahav}, {Laigle}, {La Marca}, {La Marle}, {Lamine}, {Lam}, {Lan{\c{c}}on}, {Landt}, {Langer}, {Lapi}, {Larcheveque}, {Larsen}, {Lattanzi}, {Laudisio}, {Laugier}, {Laureijs}, {Lavaux}, {Lawrenson}, {Lazanu}, {Lazeyras}, {Le Boulc'h}, {Le Brun}, {Le Brun}, {Leclercq}, {Lee}, {Le Graet}, {Legrand}, {Leirvik}, {Le Jeune}, {Lembo}, {Le
  Mignant}, {Lepinzan}, {Lepori}, {Lesci}, {Lesgourgues}, {Leuzzi}, {Levi}, {Liaudat}, {Libet}, {Liebing}, {Ligori}, {Lilje}, {Lin}, {Linde}, {Linder}, {Lindholm}, {Linke}, {Li}, {Liu}, {Lloro}, {Lobo}, {Lodieu}, {Lombardi}, {Lombriser}, {Lonare}, {Longo}, {L{\'o}pez-Caniego}, {Lopez Lopez}, {Alvarez}, {Loureiro}, {Loveday}, {Lusso}, {Macias-Perez}, {Maciaszek}, {Magliocchetti}, {Magnard}, {Magnier}, {Magro}, {Mahler}, {Mainetti}, {Maino}, {Maiorano}, {Maiorano}, {Malavasi}, {Mamon}, {Mancini}, {Mandelbaum}, {Manera}, {Manj{\'o}n-Garc{\'\i}a}, {Mannucci}, {Mansutti}, {Manteiga Outeiro}, {Maoli}, {Maraston}, {Marcin}, {Marcos-Arenal}, {Margalef-Bentabol}, {Marggraf}, {Marinucci}, {Marinucci}, {Markovic}, {Marleau}, {Marpaud}, {Martignac}, {Mart{\'\i}n-Fleitas}, {Martin-Moruno}, {Martin}, {Martinelli}, {Martinet}, {Martin}, {Martins}, {Marulli}, {Massari}, {Massey}, {Masters}, {Matarrese}, {Matsuoka}, {Matthew}, {Maughan}, {Mauri}, {Maurin}, {Maurogordato}, {McCarthy}, {McConnachie}, {McCracken}, {McDonald},
  {McEwen}, {McPartland}, {Medinaceli}, {Mehta}, {Mei}, {Melchior}, {Melin}, {M{\'e}nard}, {Mendes}, {Mendez-Abreu}, {Meneghetti}, {Mercurio}, {Merlin}, {Metcalf}, {Meylan}, {Migliaccio}, {Mignoli}, {Miller}, {Miluzio}, {Milvang-Jensen}, {Mimoso}, {Miquel}, {Miyatake}, {Mobasher}, {Mohr}, {Monaco}, {Mongui{\'o}}, {Montoro}, {Mora}, {Moradinezhad Dizgah}, {Moresco}, {Moretti}, {Morgante}, {Morisset}, {Moriya}, {Morris}, {Mortlock}, {Moscardini}, {Mota}, {Moustakas}, {Moutard}, {M{\"u}ller}, {Munari}, {Murphree}, {Murray}, {Murray}, {Musi}, {Nadathur}, {Nagam}, {Nagao}, {Naidoo}, {Nakajima}, {Nally}, {Natoli}, {Navarro-Alsina}, {Navarro Girones}, {Neissner}, {Nersesian}, {Nesseris}, {Nguyen-Kim}, {Nicastro}, {Nichol}, {Nielbock}, {Niemi}, {Nieto}, {Nilsson}, {Noller}, {Norberg}, {Nourizonoz}, {Ntelis}, {Nucita}, {Nugent}, {Nunes}, {Nutma}, {Ocampo}, {Odier}, {Oesch}, {Oguri}, {Magalhaes Oliveira}, {Onoue}, {Oosterbroek}, {Oppizzi}, {Ordenovic}, {Osato}, {Pacaud}, {Pace}, {Padilla}, {Paech}, {Pagano}, {Page},
  {Palazzi}, {Paltani}, {Pamuk}, {Pandolfi}, {Paoletti}, {Paolillo}, {Papaderos}, {Pardede}, {Parimbelli}, {Parmar}, {Partmann}, {Pasian}, {Passalacqua}, {Paterson}, {Patrizii}, {Pattison}, {Paulino-Afonso}, {Paviot}, {Peacock}, {Pearce}, {Pedersen}, {Peel}, {Peletier}, {Pellejero Ibanez}, {Pello}, {Penny}, {Percival}, {Perez-Garrido}, {Perotto}, {Pettorino}, {Pezzotta}, {Pezzuto}, {Philippon}, {Piersanti}, {Pietroni}, {Piga}, {Pilo}, {Pires}, {Pisani}, {Pizzella}, {Pizzuti}, {Plana}, {Polenta}, {Pollack}, {Poncet}, {P{\"o}ntinen}, {Pool}, {Popa}, {Popa}, {Popp}, {Porciani}, {Porth}, {Potter}, {Poulain}, {Pourtsidou}, {Pozzetti}, {Prandoni}, {Pratt}, {Prezelus}, {Prieto}, {Pugno}, {Quai}, {Quilley}, {Racca}, {Raccanelli}, {R{\'a}cz}, {Radinovi{\'c}}, {Radovich}, {Ragagnin}, {Ragnit}, {Raison}, {Ramos-Chernenko}, {Ranc}, {Raylet}, {Rebolo}, {Refregier}, {Reimberg}, {Reiprich}, {Renk}, {Renzi}, {Retre}, {Revaz}, {Reyl{\'e}}, {Reynolds}, {Rhodes}, {Ricci}, {Ricci}, {Riccio}, {Ricken}, {Rissanen}, {Risso}, {Rix},
  {Robin}, {Rocca-Volmerange}, {Rocci}, {Rodenhuis}, {Rodighiero}, {Rodriguez Monroy}, {Rollins}, {Romanello}, {Roman}, {Romelli}, {Romero-Gomez}, {Roncarelli}, {Rosati}, {Rosset}, {Rossetti}, {Roster}, {Rottgering}, {Rozas-Fern{\'a}ndez}, {Ruane}, {Rubino-Martin}, {Rudolph}, {Ruppin}, {Rusholme}, {Sacquegna}, {S{\'a}ez-Casares}, {Saga}, {Saglia}, {Sahl{\'e}n}, {Saifollahi}, {Sakr}, {Salvalaggio}, {Salvaterra}, {Salvati}, {Salvato}, {Salvignol}, {S{\'a}nchez}, {Sanchez}, {Sanders}, {Sapone}, {Saponara}, {Sarpa}, {Sarron}, {Sartori}, {Sassolas}, {Sauniere}, {Sauvage}, {Sawicki}, {Scaramella}, {Scarlata}, {Scharr{\'e}}, {Schaye}, {Schewtschenko}, {Schindler}, {Schinnerer}, {Schirmer}, {Schmidt}, {Schmidt}, {Schmidt}, {Schneider}, {Schneider}, {Schneider}, {Sch{\"o}neberg}, {Schrabback}, {Schultheis}, {Schulz}, {Schwartz}, {Sciotti}, {Scodeggio}, {Scognamiglio}, {Scott}, {Scottez}, {Secroun}, {Sefusatti}, {Seidel}, {Seiffert}, {Sellentin}, {Selwood}, {Semboloni}, {Sereno}, {Serjeant}, {Serrano}, {Shankar},
  {Sharples}, {Short}, {Shulevski}, {Shuntov}, {Sias}, {Sikkema}, {Silvestri}, {Simon}, {Sirignano}, {Sirri}, {Skottfelt}, {Slezak}, {Sluse}, {Smith}, {Smith}, {Smith}, {Smit}, {Soldano}, {Solheim}, {Sorce}, {Sorrenti}, {Soubrie}, {Spinoglio}, {Spurio Mancini}, {Stadel}, {Stagnaro}, {Stanco}, {Stanford}, {Starck}, {Stassi}, {Steinwagner}, {Stern}, {Stone}, {Strada}, {Strafella}, {Stramaccioni}, {Surace}, {Sureau}, {Suyu}, {Swindells}, {Szafraniec}, {Szapudi}, {Taamoli}, {Talia}, {Tallada-Cresp{\'\i}}, {Tanidis}, {Tao}, {Tarr{\'\i}o}, {Tavagnacco}, {Taylor}, {Taylor}, {Taylor}, {Teixeira}, {Tenti}, {Teodoro Idiago}, {Teplitz}, {Tereno}, {Tessore}, {Testa}, {Testera}, {Tewes}, {Teyssier}, {Theret}, {Thizy}, {Thomas}, {Toba}, {Toft}, {Toledo-Moreo}, {Tolstoy}, {Tommasi}, {Torbaniuk}, {Torradeflot}, {Tortora}, {Tosi}, {Tosti}, {Trifoglio}, {Troja}, {Trombetti}, {Tronconi}, {Tsedrik}, {Tsyganov}, {Tucci}, {Tutusaus}, {Uhlemann}, {Ulivi}, {Urbano}, {Vacher}, {Vaillon}, {Valdes}, {Valentijn}, {Valenziano},
  {Valieri}, {Valiviita}, {Van den Broeck}, {Vassallo}, {Vavrek}, {Venemans}, {Venhola}, {Ventura}, {Verdoes Kleijn}, {Vergani}, {Verma}, {Vernizzi}, {Veropalumbo}, {Verza}, {Vescovi}, {Vibert}, {Viel}, {Vielzeuf}, {Viglione}, {Viitanen}, {Villaescusa-Navarro}, {Vinciguerra}, {Visticot}, {Voggel}, {von Wietersheim-Kramsta}, {Vriend}, {Wachter}, {Walmsley}, {Walth}, {Walton}, {Walton}, {Wander}, {Wang}, {Wang}, {Weaver}, {Weller}, {Whalen}, {Wiesmann}, {Wilde}, {Williams}, {Winther}, {Wittje}, {Wong}, {Wright}, {Yankelevich}, {Yeung}, {Youles}, {Yung}, {Zacchei}, {Zalesky}, {Zamorani}, {Zamorano Vitorelli}, {Zanoni Marc}, {Zennaro}, {Zerbi}, {Zinchenko}, {Zoubian}, {Zucca}, \& {Zumalacarregui}}]{Euclid2024}
{Euclid Collaboration}, {Mellier}, Y., {Abdurro'uf}, {et~al.} 2024, arXiv e-prints, arXiv:2405.13491

\bibitem[{{Fian} {et~al.}(2021){Fian}, {Mediavilla}, {Motta}, {Jim{\'e}nez-Vicente}, {Mu{\~n}oz}, {Chelouche}, \& {Hanslmeier}}]{Fian2021}
{Fian}, C., {Mediavilla}, E., {Motta}, V., {et~al.} 2021, \aap, 653, A109

\bibitem[{{Filipp} {et~al.}(2023){Filipp}, {Shu}, {Pakmor}, {Suyu}, \& {Huang}}]{Filipp2023}
{Filipp}, A., {Shu}, Y., {Pakmor}, R., {Suyu}, S.~H., \& {Huang}, X. 2023, \aap, 677, A113

\bibitem[{{Findlay} {et~al.}(2018){Findlay}, {Prochaska}, {Hennawi}, {Fumagalli}, {Myers}, {Bartle}, {Chehade}, {DiPompeo}, {Shanks}, {Lau}, \& {Rubin}}]{Findlay2018}
{Findlay}, J.~R., {Prochaska}, J.~X., {Hennawi}, J.~F., {et~al.} 2018, \apjs, 236, 44

\bibitem[{{Gaia Collaboration} {et~al.}(2018){Gaia Collaboration}, {Brown}, {Vallenari}, {Prusti}, {de Bruijne}, {Babusiaux}, {Bailer-Jones}, {Biermann}, {Evans}, {Eyer}, {Jansen}, {Jordi}, {Klioner}, {Lammers}, {Lindegren}, {Luri}, {Mignard}, {Panem}, {Pourbaix}, {Randich}, {Sartoretti}, {Siddiqui}, {Soubiran}, {van Leeuwen}, {Walton}, {Arenou}, {Bastian}, {Cropper}, {Drimmel}, {Katz}, {Lattanzi}, {Bakker}, {Cacciari}, {Casta{\~n}eda}, {Chaoul}, {Cheek}, {De Angeli}, {Fabricius}, {Guerra}, {Holl}, {Masana}, {Messineo}, {Mowlavi}, {Nienartowicz}, {Panuzzo}, {Portell}, {Riello}, {Seabroke}, {Tanga}, {Th{\'e}venin}, {Gracia-Abril}, {Comoretto}, {Garcia-Reinaldos}, {Teyssier}, {Altmann}, {Andrae}, {Audard}, {Bellas-Velidis}, {Benson}, {Berthier}, {Blomme}, {Burgess}, {Busso}, {Carry}, {Cellino}, {Clementini}, {Clotet}, {Creevey}, {Davidson}, {De Ridder}, {Delchambre}, {Dell'Oro}, {Ducourant}, {Fern{\'a}ndez-Hern{\'a}ndez}, {Fouesneau}, {Fr{\'e}mat}, {Galluccio}, {Garc{\'\i}a-Torres},
  {Gonz{\'a}lez-N{\'u}{\~n}ez}, {Gonz{\'a}lez-Vidal}, {Gosset}, {Guy}, {Halbwachs}, {Hambly}, {Harrison}, {Hern{\'a}ndez}, {Hestroffer}, {Hodgkin}, {Hutton}, {Jasniewicz}, {Jean-Antoine-Piccolo}, {Jordan}, {Korn}, {Krone-Martins}, {Lanzafame}, {Lebzelter}, {L{\"o}ffler}, {Manteiga}, {Marrese}, {Mart{\'\i}n-Fleitas}, {Moitinho}, {Mora}, {Muinonen}, {Osinde}, {Pancino}, {Pauwels}, {Petit}, {Recio-Blanco}, {Richards}, {Rimoldini}, {Robin}, {Sarro}, {Siopis}, {Smith}, {Sozzetti}, {S{\"u}veges}, {Torra}, {van Reeven}, {Abbas}, {Abreu Aramburu}, {Accart}, {Aerts}, {Altavilla}, {{\'A}lvarez}, {Alvarez}, {Alves}, {Anderson}, {Andrei}, {Anglada Varela}, {Antiche}, {Antoja}, {Arcay}, {Astraatmadja}, {Bach}, {Baker}, {Balaguer-N{\'u}{\~n}ez}, {Balm}, {Barache}, {Barata}, {Barbato}, {Barblan}, {Barklem}, {Barrado}, {Barros}, {Barstow}, {Bartholom{\'e} Mu{\~n}oz}, {Bassilana}, {Becciani}, {Bellazzini}, {Berihuete}, {Bertone}, {Bianchi}, {Bienaym{\'e}}, {Blanco-Cuaresma}, {Boch}, {Boeche}, {Bombrun}, {Borrachero},
  {Bossini}, {Bouquillon}, {Bourda}, {Bragaglia}, {Bramante}, {Breddels}, {Bressan}, {Brouillet}, {Br{\"u}semeister}, {Brugaletta}, {Bucciarelli}, {Burlacu}, {Busonero}, {Butkevich}, {Buzzi}, {Caffau}, {Cancelliere}, {Cannizzaro}, {Cantat-Gaudin}, {Carballo}, {Carlucci}, {Carrasco}, {Casamiquela}, {Castellani}, {Castro-Ginard}, {Charlot}, {Chemin}, {Chiavassa}, {Cocozza}, {Costigan}, {Cowell}, {Crifo}, {Crosta}, {Crowley}, {Cuypers}, {Dafonte}, {Damerdji}, {Dapergolas}, {David}, {David}, {de Laverny}, {De Luise}, {De March}, {de Martino}, {de Souza}, {de Torres}, {Debosscher}, {del Pozo}, {Delbo}, {Delgado}, {Delgado}, {Di Matteo}, {Diakite}, {Diener}, {Distefano}, {Dolding}, {Drazinos}, {Dur{\'a}n}, {Edvardsson}, {Enke}, {Eriksson}, {Esquej}, {Eynard Bontemps}, {Fabre}, {Fabrizio}, {Faigler}, {Falc{\~a}o}, {Farr{\`a}s Casas}, {Federici}, {Fedorets}, {Fernique}, {Figueras}, {Filippi}, {Findeisen}, {Fonti}, {Fraile}, {Fraser}, {Fr{\'e}zouls}, {Gai}, {Galleti}, {Garabato}, {Garc{\'\i}a-Sedano}, {Garofalo},
  {Garralda}, {Gavel}, {Gavras}, {Gerssen}, {Geyer}, {Giacobbe}, {Gilmore}, {Girona}, {Giuffrida}, {Glass}, {Gomes}, {Granvik}, {Gueguen}, {Guerrier}, {Guiraud}, {Guti{\'e}rrez-S{\'a}nchez}, {Haigron}, {Hatzidimitriou}, {Hauser}, {Haywood}, {Heiter}, {Helmi}, {Heu}, {Hilger}, {Hobbs}, {Hofmann}, {Holland}, {Huckle}, {Hypki}, {Icardi}, {Jan{\ss}en}, {Jevardat de Fombelle}, {Jonker}, {Juh{\'a}sz}, {Julbe}, {Karampelas}, {Kewley}, {Klar}, {Kochoska}, {Kohley}, {Kolenberg}, {Kontizas}, {Kontizas}, {Koposov}, {Kordopatis}, {Kostrzewa-Rutkowska}, {Koubsky}, {Lambert}, {Lanza}, {Lasne}, {Lavigne}, {Le Fustec}, {Le Poncin-Lafitte}, {Lebreton}, {Leccia}, {Leclerc}, {Lecoeur-Taibi}, {Lenhardt}, {Leroux}, {Liao}, {Licata}, {Lindstr{\o}m}, {Lister}, {Livanou}, {Lobel}, {L{\'o}pez}, {Managau}, {Mann}, {Mantelet}, {Marchal}, {Marchant}, {Marconi}, {Marinoni}, {Marschalk{\'o}}, {Marshall}, {Martino}, {Marton}, {Mary}, {Massari}, {Matijevi{\v{c}}}, {Mazeh}, {McMillan}, {Messina}, {Michalik}, {Millar}, {Molina}, {Molinaro},
  {Moln{\'a}r}, {Montegriffo}, {Mor}, {Morbidelli}, {Morel}, {Morris}, {Mulone}, {Muraveva}, {Musella}, {Nelemans}, {Nicastro}, {Noval}, {O'Mullane}, {Ord{\'e}novic}, {Ord{\'o}{\~n}ez-Blanco}, {Osborne}, {Pagani}, {Pagano}, {Pailler}, {Palacin}, {Palaversa}, {Panahi}, {Pawlak}, {Piersimoni}, {Pineau}, {Plachy}, {Plum}, {Poggio}, {Poujoulet}, {Pr{\v{s}}a}, {Pulone}, {Racero}, {Ragaini}, {Rambaux}, {Ramos-Lerate}, {Regibo}, {Reyl{\'e}}, {Riclet}, {Ripepi}, {Riva}, {Rivard}, {Rixon}, {Roegiers}, {Roelens}, {Romero-G{\'o}mez}, {Rowell}, {Royer}, {Ruiz-Dern}, {Sadowski}, {Sagrist{\`a} Sell{\'e}s}, {Sahlmann}, {Salgado}, {Salguero}, {Sanna}, {Santana-Ros}, {Sarasso}, {Savietto}, {Schultheis}, {Sciacca}, {Segol}, {Segovia}, {S{\'e}gransan}, {Shih}, {Siltala}, {Silva}, {Smart}, {Smith}, {Solano}, {Solitro}, {Sordo}, {Soria Nieto}, {Souchay}, {Spagna}, {Spoto}, {Stampa}, {Steele}, {Steidelm{\"u}ller}, {Stephenson}, {Stoev}, {Suess}, {Surdej}, {Szabados}, {Szegedi-Elek}, {Tapiador}, {Taris}, {Tauran}, {Taylor},
  {Teixeira}, {Terrett}, {Teyssandier}, {Thuillot}, {Titarenko}, {Torra Clotet}, {Turon}, {Ulla}, {Utrilla}, {Uzzi}, {Vaillant}, {Valentini}, {Valette}, {van Elteren}, {Van Hemelryck}, {van Leeuwen}, {Vaschetto}, {Vecchiato}, {Veljanoski}, {Viala}, {Vicente}, {Vogt}, {von Essen}, {Voss}, {Votruba}, {Voutsinas}, {Walmsley}, {Weiler}, {Wertz}, {Wevers}, {Wyrzykowski}, {Yoldas}, {{\v{Z}}erjal}, {Ziaeepour}, {Zorec}, {Zschocke}, {Zucker}, {Zurbach}, \& {Zwitter}}]{GaiaCollaboration2018}
{Gaia Collaboration}, {Brown}, A.~G.~A., {Vallenari}, A., {et~al.} 2018, \aap, 616, A1

\bibitem[{Guerras {et~al.}(2013)Guerras, Mediavilla, Jimenez-Vicente, Kochanek, Muñoz, Falco, Motta, \& Rojas}]{Guerras2013}
Guerras, E., Mediavilla, E., Jimenez-Vicente, J., {et~al.} 2013, The Astrophysical Journal, 778, 123

\bibitem[{{Harris} {et~al.}(2020){Harris}, {Millman}, {van der Walt}, {Gommers}, {Virtanen}, {Cournapeau}, {Wieser}, {Taylor}, {Berg}, {Smith}, {Kern}, {Picus}, {Hoyer}, {van Kerkwijk}, {Brett}, {Haldane}, {del R{\'\i}o}, {Wiebe}, {Peterson}, {G{\'e}rard-Marchant}, {Sheppard}, {Reddy}, {Weckesser}, {Abbasi}, {Gohlke}, \& {Oliphant}}]{Harris2020}
{Harris}, C.~R., {Millman}, K.~J., {van der Walt}, S.~J., {et~al.} 2020, \nat, 585, 357

\bibitem[{{He} \& {Li}(2022)}]{He2022}
{He}, Z. \& {Li}, N. 2022, RAA, 22, 095021

\bibitem[{{He} {et~al.}(2023){He}, {Li}, {Cao}, {Li}, {Zou}, \& {Dye}}]{He2023}
{He}, Z., {Li}, N., {Cao}, X., {et~al.} 2023, \aap, 672, A123

\bibitem[{{Hennawi} {et~al.}(2010){Hennawi}, {Myers}, {Shen}, {Strauss}, {Djorgovski}, {Fan}, {Glikman}, {Mahabal}, {Martin}, {Richards}, {Schneider}, \& {Shankar}}]{Hennawi2010}
{Hennawi}, J.~F., {Myers}, A.~D., {Shen}, Y., {et~al.} 2010, \apj, 719, 1672

\bibitem[{{Hutsem{\'e}kers} \& {Sluse}(2021)}]{Hutsemekers2021}
{Hutsem{\'e}kers}, D. \& {Sluse}, D. 2021, \aap, 654, A155

\bibitem[{Kormann {et~al.}(1994)Kormann, Schneider, \& Bartelmann}]{Kormann1994}
Kormann, R., Schneider, P., \& Bartelmann, M. 1994, \aap, 284, 285

\bibitem[{{Lau} {et~al.}(2018){Lau}, {Prochaska}, \& {Hennawi}}]{Lau2018}
{Lau}, M.~W., {Prochaska}, J.~X., \& {Hennawi}, J.~F. 2018, \apj, 857, 126

\bibitem[{{Laureijs} {et~al.}(2011){Laureijs}, {Amiaux}, {Arduini}, {Augu{\`e}res}, {Brinchmann}, {Cole}, {Cropper}, {Dabin}, {Duvet}, {Ealet}, {Garilli}, {Gondoin}, {Guzzo}, {Hoar}, {Hoekstra}, {Holmes}, {Kitching}, {Maciaszek}, {Mellier}, {Pasian}, {Percival}, {Rhodes}, {Saavedra Criado}, {Sauvage}, {Scaramella}, {Valenziano}, {Warren}, {Bender}, {Castander}, {Cimatti}, {Le F{\`e}vre}, {Kurki-Suonio}, {Levi}, {Lilje}, {Meylan}, {Nichol}, {Pedersen}, {Popa}, {Rebolo Lopez}, {Rix}, {Rottgering}, {Zeilinger}, {Grupp}, {Hudelot}, {Massey}, {Meneghetti}, {Miller}, {Paltani}, {Paulin-Henriksson}, {Pires}, {Saxton}, {Schrabback}, {Seidel}, {Walsh}, {Aghanim}, {Amendola}, {Bartlett}, {Baccigalupi}, {Beaulieu}, {Benabed}, {Cuby}, {Elbaz}, {Fosalba}, {Gavazzi}, {Helmi}, {Hook}, {Irwin}, {Kneib}, {Kunz}, {Mannucci}, {Moscardini}, {Tao}, {Teyssier}, {Weller}, {Zamorani}, {Zapatero Osorio}, {Boulade}, {Foumond}, {Di Giorgio}, {Guttridge}, {James}, {Kemp}, {Martignac}, {Spencer}, {Walton}, {Bl{\"u}mchen}, {Bonoli},
  {Bortoletto}, {Cerna}, {Corcione}, {Fabron}, {Jahnke}, {Ligori}, {Madrid}, {Martin}, {Morgante}, {Pamplona}, {Prieto}, {Riva}, {Toledo}, {Trifoglio}, {Zerbi}, {Abdalla}, {Douspis}, {Grenet}, {Borgani}, {Bouwens}, {Courbin}, {Delouis}, {Dubath}, {Fontana}, {Frailis}, {Grazian}, {Koppenh{\"o}fer}, {Mansutti}, {Melchior}, {Mignoli}, {Mohr}, {Neissner}, {Noddle}, {Poncet}, {Scodeggio}, {Serrano}, {Shane}, {Starck}, {Surace}, {Taylor}, {Verdoes-Kleijn}, {Vuerli}, {Williams}, {Zacchei}, {Altieri}, {Escudero Sanz}, {Kohley}, {Oosterbroek}, {Astier}, {Bacon}, {Bardelli}, {Baugh}, {Bellagamba}, {Benoist}, {Bianchi}, {Biviano}, {Branchini}, {Carbone}, {Cardone}, {Clements}, {Colombi}, {Conselice}, {Cresci}, {Deacon}, {Dunlop}, {Fedeli}, {Fontanot}, {Franzetti}, {Giocoli}, {Garcia-Bellido}, {Gow}, {Heavens}, {Hewett}, {Heymans}, {Holland}, {Huang}, {Ilbert}, {Joachimi}, {Jennins}, {Kerins}, {Kiessling}, {Kirk}, {Kotak}, {Krause}, {Lahav}, {van Leeuwen}, {Lesgourgues}, {Lombardi}, {Magliocchetti}, {Maguire},
  {Majerotto}, {Maoli}, {Marulli}, {Maurogordato}, {McCracken}, {McLure}, {Melchiorri}, {Merson}, {Moresco}, {Nonino}, {Norberg}, {Peacock}, {Pello}, {Penny}, {Pettorino}, {Di Porto}, {Pozzetti}, {Quercellini}, {Radovich}, {Rassat}, {Roche}, {Ronayette}, {Rossetti}, {Sartoris}, {Schneider}, {Semboloni}, {Serjeant}, {Simpson}, {Skordis}, {Smadja}, {Smartt}, {Spano}, {Spiro}, {Sullivan}, {Tilquin}, {Trotta}, {Verde}, {Wang}, {Williger}, {Zhao}, {Zoubian}, \& {Zucca}}]{Euclid-intro}
{Laureijs}, R., {Amiaux}, J., {Arduini}, S., {et~al.} 2011, arXiv e-prints, arXiv:1110.3193

\bibitem[{{Lemon} {et~al.}(2022){Lemon}, {Anguita}, {Auger-Williams}, {Courbin}, {Galan}, {McMahon}, {Neira}, {Oguri}, {Schechter}, {Shajib}, {Treu}, {Agnello}, \& {Spiniello}}]{lemon2022}
{Lemon}, C., {Anguita}, T., {Auger-Williams}, M.~W., {et~al.} 2022, \mnras

\bibitem[{{Li} {et~al.}(2021){Li}, {Napolitano}, {Spiniello}, {Tortora}, {Kuijken}, {Koopmans}, {Schneider}, {Getman}, {Xie}, {Long}, {Shu}, {Vernardos}, {Huang}, {Covone}, {Dvornik}, {Heymans}, {Hildebrandt}, {Radovich}, \& {Wright}}]{Li2021}
{Li}, R., {Napolitano}, N.~R., {Spiniello}, C., {et~al.} 2021, \apj, 923, 16

\bibitem[{Liao {et~al.}(2019)Liao, Shafieloo, Keeley, \& Linder}]{Liao2019}
Liao, K., Shafieloo, A., Keeley, R.~E., \& Linder, E.~V. 2019, \apj, 886, L23

\bibitem[{{Marshall} {et~al.}(2016){Marshall}, {Verma}, {More}, {Davis}, {More}, {Kapadia}, {Parrish}, {Snyder}, {Wilcox}, {Baeten}, {Macmillan}, {Cornen}, {Baumer}, {Simpson}, {Lintott}, {Miller}, {Paget}, {Simpson}, {Smith}, {K{\"u}ng}, {Saha}, \& {Collett}}]{Marshall2016}
{Marshall}, P.~J., {Verma}, A., {More}, A., {et~al.} 2016, \mnras, 455, 1171

\bibitem[{{Martin} {et~al.}(2018){Martin}, {Kaviraj}, {Devriendt}, {Dubois}, \& {Pichon}}]{Martin2018}
{Martin}, G., {Kaviraj}, S., {Devriendt}, J.~E.~G., {Dubois}, Y., \& {Pichon}, C. 2018, \mnras, 480, 2266

\bibitem[{McCully {et~al.}(2018)McCully, Crawford, Kovacs, Tollerud, Betts, Bradley, Craig, Turner, Streicher, Sipocz, Robitaille, \& Deil}]{McCully2018}
McCully, C., Crawford, S., Kovacs, G., {et~al.} 2018, astropy/astroscrappy: v1.0.5 Zenodo Release

\bibitem[{{Motta} {et~al.}(2012){Motta}, {Mediavilla}, {Falco}, \& {Mu{\~n}oz}}]{Motta2012}
{Motta}, V., {Mediavilla}, E., {Falco}, E., \& {Mu{\~n}oz}, J.~A. 2012, \apj, 755, 82

\bibitem[{{Oguri} \& {Marshall}(2010)}]{Oguri2010}
{Oguri}, M. \& {Marshall}, P.~J. 2010, \mnras, 405, 2579

\bibitem[{{Oguri} {et~al.}(2014){Oguri}, {Rusu}, \& {Falco}}]{Oguri2014}
{Oguri}, M., {Rusu}, C.~E., \& {Falco}, E.~E. 2014, \mnras, 439, 2494

\bibitem[{{Oke} \& {Gunn}(1982)}]{Oke1982}
{Oke}, J.~B. \& {Gunn}, J.~E. 1982, \pasp, 94, 586

\bibitem[{{Roedig} {et~al.}(2014){Roedig}, {Krolik}, \& {Miller}}]{Roedig2014}
{Roedig}, C., {Krolik}, J.~H., \& {Miller}, M.~C. 2014, \apj, 785, 115

\bibitem[{{Romero} {et~al.}(2016){Romero}, {Vila}, \& {P{\'e}rez}}]{Romero2016}
{Romero}, G.~E., {Vila}, G.~S., \& {P{\'e}rez}, D. 2016, \aap, 588, A125

\bibitem[{{Shajib} {et~al.}(2018){Shajib}, {Treu}, \& {Agnello}}]{Shajib2018}
{Shajib}, A.~J., {Treu}, T., \& {Agnello}, A. 2018, \mnras, 473, 210

\bibitem[{{Shu} {et~al.}(2015){Shu}, {Bolton}, {Brownstein}, {Montero-Dorta}, {Koopmans}, {Treu}, {Gavazzi}, {Auger}, {Czoske}, {Marshall}, \& {Moustakas}}]{Shu2015}
{Shu}, Y., {Bolton}, A.~S., {Brownstein}, J.~R., {et~al.} 2015, \apj, 803, 71

\bibitem[{{Shu} {et~al.}(2022){Shu}, {Ca{\~n}ameras}, {Schuldt}, {Suyu}, {Taubenberger}, {Inoue}, \& {Jaelani}}]{Shu2022}
{Shu}, Y., {Ca{\~n}ameras}, R., {Schuldt}, S., {et~al.} 2022, \aap, 662, A4

\bibitem[{{Sluse} {et~al.}(2011){Sluse}, {Schmidt}, {Courbin}, {Hutsem{\'e}kers}, {Meylan}, {Eigenbrod}, {Anguita}, {Agol}, \& {Wambsganss}}]{Sluse2011}
{Sluse}, D., {Schmidt}, R., {Courbin}, F., {et~al.} 2011, \aap, 528, A100

\bibitem[{{Sonnenfeld} \& {Cautun}(2021)}]{Sonnenfeld2021}
{Sonnenfeld}, A. \& {Cautun}, M. 2021, \aap, 651, A18

\bibitem[{{Sonnenfeld} {et~al.}(2013){Sonnenfeld}, {Treu}, {Gavazzi}, {Suyu}, {Marshall}, {Auger}, \& {Nipoti}}]{Sonnenfeld13}
{Sonnenfeld}, A., {Treu}, T., {Gavazzi}, R., {et~al.} 2013, \apj, 777, 98

\bibitem[{{Suyu} {et~al.}(2014){Suyu}, {Treu}, {Hilbert}, {Sonnenfeld}, {Auger}, {Blandford}, {Collett}, {Courbin}, {Fassnacht}, {Koopmans}, {Marshall}, {Meylan}, {Spiniello}, \& {Tewes}}]{Suyu2014}
{Suyu}, S.~H., {Treu}, T., {Hilbert}, S., {et~al.} 2014, \apjl, 788, L35

\bibitem[{{The Dark Energy Survey Collaboration}(2005)}]{DES2005}
{The Dark Energy Survey Collaboration}. 2005, arXiv e-prints, astro

\bibitem[{{W}es {M}c{K}inney(2010)}]{McKinney2010}
{W}es {M}c{K}inney. 2010, in {P}roceedings of the 9th {P}ython in {S}cience {C}onference, ed. {S}t\'efan van~der {W}alt \& {J}arrod {M}illman, 56 -- 61

\bibitem[{{Wong} {et~al.}(2020){Wong}, {Suyu}, {Chen}, {Rusu}, {Millon}, {Sluse}, {Bonvin}, {Fassnacht}, {Taubenberger}, {Auger}, {Birrer}, {Chan}, {Courbin}, {Hilbert}, {Tihhonova}, {Treu}, {Agnello}, {Ding}, {Jee}, {Komatsu}, {Shajib}, {Sonnenfeld}, {Blandford}, {Koopmans}, {Marshall}, \& {Meylan}}]{Wong2020}
{Wong}, K.~C., {Suyu}, S.~H., {Chen}, G. C.~F., {et~al.} 2020, \mnras, 498, 1420

\bibitem[{{Zhu} {et~al.}(2023){Zhu}, {Shu}, {Yuan}, {Fu}, {Gao}, {Wu}, {He}, {Liao}, {Li}, {Er}, \& {Hu}}]{Zhu2023}
{Zhu}, S., {Shu}, Y., {Yuan}, H., {et~al.} 2023, Research in Astronomy and Astrophysics, 23, 035001

\end{thebibliography}

\begin{appendix} 
\section{Light and mass modelling}
\label{sec:lens_modelling}
We use the two-Moffat profile $I_{P,i}$ ($i$ presents the $i$-th image) to describe the light distributions of the quasars in each band, which can be considered as PSFs in DESI-LS images. $I_{P,i}$ is characterized by the following parameters: the width parameters ($\alpha_1$, $\alpha_2$), shape parameters ($\beta_1$, $\beta_2$), position angles ($\phi_{m,1}$, $\phi_{m,2}$), offsets between the centres of the first and second Moffat profiles ($\delta x_{m}$, $\delta y_{m}$), the amplitude ratio between the first and second Moffat profiles ($f$), the amplitude of the first Moffat profile ($A_{m}^i$), and the coordinates of the centre of the first Moffat profile ($x_{m}^i$, $y_{m}^i$).
We determine the parameters $\alpha_1$, $\alpha_2$, $\beta_1$, $\beta_2$, $\phi_{m,1}$, $\phi_{m,2}$, $\delta x_{m}$, $\delta y_{m}$, and $f$ through star fitting using a nearby star close to the system of interest, and leaving $A_{m}^i$, $x_{m}^i$, and $y_{m}^i$ to be fitted in later procedures.
We employ a single Sérsic profile along with $i$ PSFs to model the light distribution of lensed quasar systems. The model can be expressed as:
\begin{equation}
    I_{model}=I_s+\Sigma_{i=1}^{nimg} I_{P,i}
\end{equation}
The Sérsic profile ($I_s$) is characterized by seven parameters: Sérsic index ($n_s$), half-light radius ($r_s$), position angle ($\phi_s$), axis-ratio ($q_s$), amplitude ($A_s$) and the coordinates of the lens centre ($x_{lens}$, $y_{lens}$).
Initially, we fit the band exhibiting the most prominent lensing galaxy signal (as indicated by the residual image of \LIS) to minimise $A_{m}^i$, $x_{m}^i$, and $y_{m}^i$ of the PSF model and $x_{lens}$, $y_{lens}$ of Sersic model use the following likelihood function. 
\begin{equation}
\label{eq:light_likelihood}
    \textrm{log} p(I_{data}|I_{model})=\frac{(I_{data}-I_{model})^2}{2\sigma_{bkg}^2},
\end{equation}
where $ \sigma_{bkg} $ denotes the background noise.
Subsequently, we fit the other bands using the same procedure. In those fittings, we fix $x_{m}^i$, $y_{m}^i$, $x_{lens}$, and $y_{lens}$ according to the results obtained from the first band. The other parameters ($A_{m}^i$, $n_s$, $r_s$, $\phi_s$, $q_s$, and $A_s$) were allowed to be differed among each bands.
We adopt a SIE as our lens mass model since it has been shown to align well with observations across numerous studies \citep[see e.g.][]{Bolton12, Sonnenfeld13}. 
During the lens modelling process, we assume that the SIE halo shares the same centre as the Sérsic profile previously determined ($x_{lens}$, $y_{lens}$). Here, to avoid the effect of microlensing and AGN light-variation, we only fit the positions of multiple images,  which are established during the light modelling phase. The parameters adjusted in mass modelling include: Einstein radius ($ \theta_E $), axis ratio and position angle of SIE ($q_{SIE}$, $\phi_{SIE}$), and source position ($ x_{src} $, $ y_{src} $). Note that $q_{SIE}$ and $\phi_{SIE}$ were constrained within a range centered on $q_s$ and $\phi_s$, which were determined during the light modeling stage based on the results from the band with the highest signal-to-noise ratio for the lensing galaxy. To be specific, $q_{SIE}$ was allowed for a $\pm 30\%$ variation compared to $q_{s}$, while $\phi_{SIE}$ was allowed for a $\pm 30^\circ$ variation compared to $\phi_{s}$. Totally, we used four inputted parameters ($x_{m}^1,x_{m}^2,y_{m}^1,y_{m}^2$) to constrain five parameters ($ \theta_E $, $ x_{src} $, $ y_{src} $, $q_{SIE}$, and $\phi_{SIE}$), and gave $q_{SIE}$ and $\phi_{SIE}$ priors.
The likelihood function used in SIE fitting is that:
\begin{equation}
\textrm{log} p(\vec{r}_{i}|\vec{r}_i^{P})=\sum_{i} \frac{\left|\vec{r}_{i}-\vec{r}_i^{P}\right|^{2}}{\sigma_{p,i}^2}
\end{equation}
with
\begin{equation}
    \vec{r}_{i}=
    \begin{pmatrix}
    x_{m}^i\\
    y_{m}^i
    \end{pmatrix}
\end{equation}
where $\sigma_{p,i}$ represents the astrometric error of $i$-th image, $\vec{r}_i^{P}$ is the position of multiple images predicted by SIE model. 
After obtaining all parameters, we can further estimate the time delay by varying the range of $ z_d $, set between 0.1 and 1.25 to encompass the $ z_d $ of the majority known lensed quasars. The time delay is defined as:
\begin{equation}
    t_d=arrival\_time({Img}_B)-arrival\_time({Img}_A),
\end{equation}
where labels A and B correspond to those in Figure \ref{fig:lens-model}. In Figure \ref{fig:td}, we illustrate the time delay versus different values of $ z_d $.
\end{appendix}
\end{document}